%
%
%
\expandafter\ifx\csname ACRES chardef at\endcsname\relax\else\endinput\fi
\expandafter\chardef\csname ACRES chardef at\endcsname=\the\catcode`\@
\expandafter\chardef\csname ACRES chardef colon\endcsname=\the\catcode`\:
\catcode`\:=11\catcode`\@=11
%
%
%
%
\def\@cres:version{ACRES version 0.996 dd 20030217 by G.M. Tuynman }
\font\@cres:tracingfont=cmr5
\def\@cres:warning#1{\immediate\write16{ACRES: #1}}
\def\@cres:newcount{\alloc@0\count\countdef\insc@unt}
%
%
%
\newread\@cres:myin
\newwrite\@cres:myout
\newif\if@cres:tracing\@cres:tracingfalse
\newif\if@cres:noteof
\def\@cres:chapsuppress{1}
\newif\if@cres:pagenos\@cres:pagenostrue
\@cres:newcount\@cres:chapdepth
\@cres:newcount\@cres:cntrdepth
\@cres:newcount\@cres:i
\@cres:newcount\@cres:rmi
\@cres:newcount\@cres:rmx
\@cres:newcount\@cres:rmc
\@cres:newcount\@cres:rmm
%
%
%
%
\def\@cres:oponchap#1#2{\expandafter#1\csname @cres:chap#2\endcsname}
\def\@cres:oponcntr#1#2{\expandafter#1\csname @cres:cntr#2\endcsname}
%
%
%
%
\def\@cres:preprm#1{\ifnum#1<1\@cres:rmi=10\@cres:rmx=0\@cres:rmc=0
  \@cres:rmm=0\else\@cres:rmi=#1\ifnum#1<4000\@cres:rmx=\@cres:rmi
  \divide\@cres:rmx by10 \multiply\@cres:rmx by10\advance\@cres:rmi
  by-\@cres:rmx \divide\@cres:rmx by10\@cres:rmc=\@cres:rmx\divide
  \@cres:rmc by10 \multiply\@cres:rmc by10\advance\@cres:rmx
  by-\@cres:rmc \divide\@cres:rmc by10\@cres:rmm=\@cres:rmc\divide
  \@cres:rmm by10 \multiply\@cres:rmm by10\advance\@cres:rmc
  by-\@cres:rmm \divide\@cres:rmm by10\else\@cres:rmx=0\@cres:rmc=0
  \@cres:rmm=#1\fi\fi\relax}
\def\@cres:prtrm{\ifcase\@cres:rmm\or m\or mm\or mmm\else\the\@cres:rmi\fi
  \ifcase \@cres:rmc \or c\or cc\or ccc\or cd\or d\or dc\or dcc\or
  dccc\or cm\fi \ifcase\@cres:rmx\or x\or xx\or xxx\or xl\or l\or lx\or
  lxx\or lxxx\or xc\fi\ifcase\@cres:rmi\or i\or ii\or iii\or iv\or v\or
  vi\or vii\or viii\or ix\or\fi}
\def\@cres:prtRM{\ifcase\@cres:rmm\or M\or MM\or MMM\else\the\@cres:rmi\fi
  \ifcase \@cres:rmc \or C\or CC\or CCC\or CD\or D\or DC\or DCC\or
  DCCC\or CM\fi \ifcase\@cres:rmx\or X\or XX\or XXX\or XL\or L\or LX\or
  LXX\or LXXX\or XC\fi\ifcase\@cres:rmi\or I\or II\or III\or IV\or V\or
  VI\or VII\or VIII\or IX\or\fi}
\def\@cres:prta#1{\ifcase#1 0\or a\or b\or c\or d\or e\or f\or g\or h\or
i\or j\or k\or l\or m\or n\or o\or p\or q\or r\or s\or t\or u\or v\or w\or
x\or y\or z\else#1\fi}
\def\@cres:prtA#1{\ifcase#1 0\or A\or B\or C\or D\or E\or F\or G\or H\or
I\or J\or K\or L\or M\or N\or O\or P\or Q\or R\or S\or T\or U\or V\or W\or
X\or Y\or Z\else#1\fi}
\def\@cres:prepprint{\ifcase\@cres:mode/\or\@cres:preprm{\@cres:number}\or
  \@cres:preprm{\@cres:number}\fi}
\def\@cres:PrintInMode{\ifcase\@cres:mode/ \@cres:number\or\@cres:prtrm\or
\@cres:prtRM\or\@cres:prta{\@cres:number}\or\@cres:prtA{\@cres:number}%
\fi}
%
%
%
%
\def\@cres:printref#1#2#3#4#5{%
\def\@cres:empty{}\def\@cres:tail{#3}%
\ifx\@cres:tail\@cres:empty 
   \def\@cres:noprint{?-.!;,}\def\@cres:sep{#2}%
   \ifx\@cres:sep\@cres:noprint\else#1\fi
\else\def\@cres:head{#1}%
    \ifx\@cres:head\@cres:empty #3%
    \else\edef\@cres:mode/{\csname @cres:chapmode1\endcsname}%
        \edef\@cres:number{\@cres:oponchap\the1}%
        \@cres:prepprint 
        \edef\@cres:chapno{\@cres:PrintInMode}%
        \ifcase\@cres:chapsuppress\relax #3%
        \or\ifx\@cres:head\@cres:chapno#3\else#1#2#3\fi
        \else#1#2#3%
        \fi 
    \fi
\fi}
%
%
%
%
\def\@cres:printpageref#1#2#3#4#5{\ifnum#4<0 
\romannumeral-#4\else#4\fi}
%
%
%
%
\def\@cres:printtitletext#1#2#3#4#5{#5}
%
%
%
%
\def\@cres:cntr#1{%
\global\@cres:oponcntr\advance#1 by1
\edef\@cres:cntrlevel{\csname @cres:cntrlev#1\endcsname}%
\edef\@cres:mode/{\csname @cres:cntrmode#1\endcsname}%
\edef\@cres:number{\@cres:oponcntr\the#1}%
  \@cres:prepprint
  \edef\@cres:counter{\@cres:PrintInMode}%
\ifnum\@cres:cntrlevel<1\xdef\@cres:newref{{}{}{\@cres:counter}{\the
                                                            \pageno}{}}%
\else\ifnum\@cres:cntrlevel<2 \xdef\@cres:newref{\@cres:counter}%
        \xdef\@cres:cntrseptr{\csname @cres:cntrsep#1\endcsname}%
     \else\edef\@cres:mode/{\csname @cres:chapmode2\endcsname}%
         \edef\@cres:number{\@cres:oponchap\the2}%
         \@cres:prepprint\xdef\@cres:newref{\@cres:PrintInMode}\@cres:i=2
         \xdef\@cres:cntrseptr{\csname @cres:chapsep2\endcsname}%
         \loop\ifnum\@cres:i<\@cres:cntrlevel\advance\@cres:i by1
             \edef\@cres:mode/{%
              \csname @cres:chapmode\the\@cres:i\endcsname}%
             \edef\@cres:number{\@cres:oponchap\the{\the\@cres:i}}%
             \@cres:prepprint\xdef\@cres:newref{\@cres:newref
             \csname @cres:chapsep\the\@cres:i\endcsname\@cres:PrintInMode}%
         \repeat
         \xdef\@cres:newref{\@cres:newref
          \csname @cres:cntrsep#1\endcsname\@cres:counter}%
     \fi
     \edef\@cres:mode/{\csname @cres:chapmode1\endcsname}%
     \edef\@cres:number{\@cres:oponchap\the1}%
     \@cres:prepprint
     \xdef\@cres:newref{{\@cres:PrintInMode}{\@cres:cntrseptr
                                        }{\@cres:newref}{\the\pageno}{}}%
\fi
\expandafter\@cres:printref\@cres:newref}
%
%
%
%
\def\newcounter#1#2#3#4{%
\expandafter
\ifx\csname ACRES initialize\endcsname\relax
   \@cres:initialize\expandafter\gdef\csname ACRES initialize\endcsname{}%
\fi
\advance\@cres:cntrdepth by1 
\@cres:oponcntr\@cres:newcount{\the\@cres:cntrdepth}%
\expandafter\xdef\csname#1\endcsname
  {\noexpand\@cres:cntr\the\@cres:cntrdepth}%
\expandafter\gdef\csname @cres:cntrsep\the\@cres:cntrdepth\endcsname{#3}%
\expandafter\xdef\csname @cres:cntrmode\the\@cres:cntrdepth\endcsname
  {\ifx#2i1\else\ifx#2I2\else\ifx#2a3\else\ifx#2A4\else0\fi\fi\fi\fi}%
\ifnum#4<\@cres:chapdepth\@cres:i=#4\else\@cres:i=\@cres:chapdepth\fi
\expandafter\xdef\csname @cres:cntrlev\the\@cres:cntrdepth\endcsname
  {\the\@cres:i}}
%
%
%
%
\def\@cres:chap#1{%
\@cres:i=#1\global\@cres:oponchap\advance#1 by1
\loop\ifnum\@cres:i<\@cres:chapdepth\advance\@cres:i by1
    \global\csname @cres:chap\the\@cres:i\endcsname=0%
\repeat
\@cres:i=0
\loop\ifnum\@cres:i<\@cres:cntrdepth\advance\@cres:i by1
    \expandafter
    \ifnum\csname @cres:cntrlev\the\@cres:i\endcsname<#1
    \else\global\csname @cres:cntr\the\@cres:i\endcsname=0
    \fi
\repeat
\ifnum#1<2 \xdef\@cres:newref{}\xdef\@cres:septr{}%
\else\edef\@cres:mode/{\csname @cres:chapmode2\endcsname}%
    \edef\@cres:number{\@cres:oponchap\the2}%
    \@cres:prepprint 
    \xdef\@cres:newref{\@cres:PrintInMode}%
    \@cres:i=2
    \xdef\@cres:septr{\csname @cres:chapsep2\endcsname}%
    \loop\ifnum\@cres:i<#1\advance\@cres:i by1
        \edef\@cres:mode/{\csname @cres:chapmode\the\@cres:i\endcsname}%
        \edef\@cres:number{\@cres:oponchap\the{\the\@cres:i}}%
        \@cres:prepprint
        \xdef\@cres:newref{\@cres:newref
            \csname @cres:chapsep\the\@cres:i\endcsname\@cres:PrintInMode}%
    \repeat
\fi
\edef\@cres:mode/{\csname @cres:chapmode1\endcsname}%
\edef\@cres:number{\@cres:oponchap\the1}%
\@cres:prepprint
\xdef\@cres:newref{{\@cres:PrintInMode}{\@cres:septr
                                         }{\@cres:newref}{\the\pageno}{}}%
\expandafter\@cres:printref\@cres:newref}
%
%
%
%
\def\newlevel#1#2#3{%
\expandafter
\ifx\csname ACRES initialize\endcsname\relax 
   \@cres:initialize\expandafter\gdef\csname ACRES initialize\endcsname{}%
\fi
\advance\@cres:chapdepth by1 
\@cres:oponchap\@cres:newcount{\the\@cres:chapdepth}%
\expandafter\xdef\csname#1\endcsname
 {\noexpand\@cres:chap\the\@cres:chapdepth}%
\expandafter\gdef\csname @cres:chapsep\the\@cres:chapdepth\endcsname{#3}%
\expandafter\xdef\csname @cres:chapmode\the\@cres:chapdepth\endcsname
    {\ifx#2i1\else\ifx#2I2\else\ifx#2a3\else\ifx#2A4\else0\fi\fi\fi\fi}}
%
%
%
%
\@cres:newcount\@cres:withoutnumber\@cres:withoutnumber=0
\def\withoutnumber{\global\advance\@cres:withoutnumber by1
   \xdef\@cres:newref{{\the\@cres:withoutnumber}{?-.!;,}{}{\the\pageno}{}}}
%
%
%
%
\def\@cres:undefined#1#2#3{%
\expandafter\ifx\csname @cres::#1\endcsname\relax#2\else#3\fi}
%
%
%
%
\def\@cres:shipout#1{{\let\the=0\edef\next{\write\@cres:myout{#1}}\next}}
%
%
%
%
\def\@cres:memorize#1#2{\expandafter\xdef\csname @cres::#1\endcsname{#2}}
%
%
%
%
\def\memorize#1=#2{%
\if@cres:tracing{\@cres:tracingfont#2=}\fi
\def\@cres:tricky{#2}%
#1%
\expandafter\@cres:strippageandtext\@cres:newref
\edef\@cres:newrefwithoutpageandtext{\@cres:refwithoutpageandtext}%
\@cres:undefined{#2}{\@cres:memorize{#2}\@cres:newref}%
  {{\edef\@cres:oldref{\csname @cres::#2\endcsname}%
    \expandafter\@cres:strippageandtext\@cres:oldref
    \ifx\@cres:refwithoutpageandtext\@cres:newrefwithoutpageandtext
    \else\@cres:warning{changed or duplicate "#2"}%
        \@cres:memorize{#2}\@cres:newref
    \fi}}%
\@cres:shipout{{#2}{\@cres:newrefwithoutpageandtext{\the\pageno}{}}}}
%
%
%
%
\def\memorizetitle#1=#2#3#4{%
\if@cres:tracing{\@cres:tracingfont#2=}\fi
\def\@cres:tricky{#2}%
#1#3#4%
\expandafter\@cres:strippageandtext\@cres:newref
\edef\@cres:newrefwithoutpageandtext{\@cres:refwithoutpageandtext}%
\@cres:undefined{#2}{\@cres:memorize{#2}\@cres:newref}%
  {{\edef\@cres:oldref{\csname @cres::#2\endcsname}%
    \expandafter\@cres:strippageandtext\@cres:oldref
    \ifx\@cres:refwithoutpageandtext\@cres:newrefwithoutpageandtext
    \else\@cres:warning{changed or duplicate "#2"}%
        \@cres:memorize{#2}\@cres:newref
    \fi}}%
\@cres:shipout{{#2}{\@cres:newrefwithoutpageandtext{\the\pageno}{#4}}}}
%
%
%
%
\def\@cres:strippageandtext#1#2#3#4#5{%
\xdef\@cres:refwithoutpageandtext{{#1}{#2}{#3}}}
%
%
%
\def\recall#1{{%
\@cres:undefined{#1}
{\@cres:warning{undefined "#1"}%
    {{\bf???}\if@cres:tracing\kern.5em {\@cres:tracingfont#1}\fi}}
{{\edef\@cres:inter{\csname @cres::#1\endcsname}%
    \expandafter\@cres:printref\@cres:inter}}}}
%
%
%
%
\def\recallpage#1{{%
\if@cres:pagenos\global\@cres:pagenosfalse
   \@cres:warning{page numbers : do not forget to typeset twice}\fi
\@cres:undefined{#1}
{\@cres:warning{undefined "#1"}%
    {{\bf???}\if@cres:tracing\kern.5em {\@cres:tracingfont#1}\fi}}
{{\edef\@cres:inter{\csname @cres::#1\endcsname}%
    \expandafter\@cres:printpageref\@cres:inter}}}}
%
%
%
%
\def\recalltitletext#1{{%
\@cres:undefined{#1}
{\@cres:warning{undefined "#1"}%
    {{\bf???}\if@cres:tracing{\kern.5em \@cres:tracingfont#1}\fi}}
{{\edef\@cres:inter{\csname @cres::#1\endcsname}%
    \expandafter\@cres:printtitletext\@cres:inter}}}}
%
%
%
%
\def\tracinglabels{\@cres:tracingtrue}
%
%
%
%
\def\previousreferences#1{{%
\openin\@cres:myin=#1.acres %
\ifeof\@cres:myin\@cres:warning{#1.acres does not exist}%
\else\read\@cres:myin to\@cres:reference
    \ifx\@cres:reference\@cres:version
        \loop\read\@cres:myin to\@cres:reference
            \ifeof\@cres:myin\@cres:noteoffalse\else\@cres:noteoftrue\fi
        \if@cres:noteof\expandafter\@cres:memorize\@cres:reference
        \repeat
    \else\@cres:warning{wrong version of ACRES}%
    \fi
\fi
\closein\@cres:myin}}
%
%
%
%
%
\def\outerlevelinclude{\def\@cres:chapsuppress{2}}
\def\outerlevelsuppress{\def\@cres:chapsuppress{1}}
\def\Outerlevelsuppress{\def\@cres:chapsuppress{0}}
%
%
%
%
\def\@cres:initialize{%
\previousreferences{\jobname}%
\openout\@cres:myout=\jobname.acres %
\@cres:shipout{\@cres:version}}
%
%
%
%
\def\IncreaseByOne#1{\expandafter\@cres:stripperplus#1}
\def\DecreaseByOne#1{\expandafter\@cres:stripperminus#1}
\def\@cres:stripperminus#1#2{%
\ifx\@cres:cntr#1\global\@cres:oponcntr\advance#2 by-1 
\else\ifx\@cres:chap#1\global\@cres:oponchap\advance#2 by-1 
    \else\@cres:warning{DecreaseByOne does not apply}\fi
\fi}
\def\@cres:stripperplus#1#2{%
\ifx\@cres:cntr#1\global\@cres:oponcntr\advance#2 by1 
\else\ifx\@cres:chap#1\global\@cres:oponchap\advance#2 by1 
    \else\@cres:warning{IncreaseByOne does not apply}\fi
\fi}
%
%
%
\catcode`\@=\csname ACRES chardef at\endcsname
\catcode`\:=\csname ACRES chardef colon\endcsname
%
%
\input amstex.tex \input amsppt.sty
\magnification=\magstep1

\long\def\dontprint#1{#1}

\newdimen\myproclaimskipamount
\myproclaimskipamount=1.5\baselineskip
\newdimen\mydefinitionskipamount
\mydefinitionskipamount=1.5\baselineskip

\newlevel{headnum}{1}{.}
{1}{.}{1}
\outerlevelinclude


\def\thm#1{{\theoremnummer\ #1}}
\def\thmm#1#2{{\memorize\theoremnummer={#1} #2}}
\def\formula#1{\tag{\memorize\theoremnummer={#1}}}
\def\recalt#1{[\recall{#1}]}
\def\recalf#1{(\recall{#1})}

\newif\iflongsign\longsigntrue

\def\altsign#1#2{\iflongsign#1\else(-1)^{#2}\fi}


\def\ad{\operatorname{ad}}
\def\Ad{\operatorname{Ad}}
\def\aglalg/{an \glalg/}
\def\aglgrp/{an \glgrp/}
\def\agmfd/{an \gmfd/}
\def\agvs/{an \gvs/}
\def\arglalg/{an \rglalg/}
\def\arvs/{an \rvs/}
\def\assmfd/{a \ssmfd/}
\def\at{\@}
\def\Aut{\operatorname{Aut}}
\def\body{{\bold B}}

\def\CA{{\Cal A}}
\def\CAC{{\CA^\CC}}
\def\cadb#1{\bold#1}
\def\CC{{\bold C}}
\def\Ci{C^\infty}
\def\Coad{\operatorname{Coad}}
\def\coad{\operatorname{coad}}
\def\comm#1#2{(-1)^{#1#2}}
\def\comme#1#2{(-1)^{\e(#1)\e(#2)}}

\def\contrf#1#2{\contrfoper(#1)#2}
\def\contrfoper{{\iota}}
\def\contrs#1#2{{\langle\mskip-5mu\langle\mskip1.5mu}
    #1\mskip1.5mu |\mskip-3mu|\mskip1.5mu #2
    {\mskip1.5mu\rangle\mskip-5mu\rangle}}
\def\contrs#1#2{{\langle\mskip1.5mu}
    #1\mskip1.5mu |\mskip1.5mu #2
    {\mskip1.5mu\rangle}}

\def\cover#1{{\Cal #1}}
\def\curv{\operatorname{curvature}}
\def\datum{\the\day/\the\month/\the\year}
\def\dom#1{[#1]}

\def\double#1{{\underline{\overline{#1}}}}
\def\e{\varepsilon}

\def\eexp{\operatorname{e}}
\def\End{\operatorname{End}}

\def\fracp#1#2{\frac{\partial #1}{\partial #2}}

\def\Gext{{\bold G}}
\def\glalg/{$\CA$-Lie algebra}
\def\glgrp/{$\CA$-Lie group}
\def\glsubgrp/{$\CA$-Lie subgroup}
\def\gmfd/{$\CA$-mani\-fold}
\def\gvs/{$\CA$-vector space}
\def\Hh{{\widehat H}}
\def\Hom{\operatorname{Hom}}
\def\HSym{\operatorname{HSymm}}
\def\ie{i.e.}
\def\im{\operatorname{im}}
\def\invol{{\frak C}}
\def\itemize{$\bullet$ }
\def\Liealg#1{{\frak#1}}
\def\Liealghh{{\widehat{\Liealg h}}}
\def\lieb#1{[\mskip3mu#1\mskip3mu]}

\def\Lied{{\Cal L}}
\def\mapob{\kern.3em} 
\def\mo{^{-1}}
\def\mub{{\bar\mu}}
\def\nerve{{\Cal N}}
\def\NN{{\bold N}}
\def\NSym{\operatorname{NSymm}}

\def\nul{\lbrace0\rbrace}
\def\Orbit{{\Cal O}}
\def\PB#1{\{#1\}}
\def\Per{\operatorname{Per}}
\def\phih{{\hat\phi}}
\def\Poisson{{\Cal P}}
\def\QEDbox{\hbox{\lower2.3pt\vbox{\hrule\hbox
   {\vrule\kern1pt\vbox{\kern1.7pt\hbox{$\scriptstyle
   QED$}\kern.6pt}\kern1pt\vrule}\hrule}}}
\def\QED{\hskip0.01em plus 40pt\null{} \null\nobreak\hfill
   \kern3pt\QEDbox} 
\def\QFS{{\Cal E}}
\def\QZO{{\Cal Q}}
\def\restricted{|}
\def\rglalg/{$\RR$-Lie algebra}
\def\rmfd/{$\RR$-manifold}
\def\rvs/{$\RR$-vector space}
\def\RR{{\bold R}}
\def\scirc{\,{\raise 0.8pt\hbox{$\scriptstyle\circ$}}\,}
\def\SS{{\bold S}}
\def\ssmfd/{symplectic \gmfd/}
\def\stress#1{{\it #1\/}}
\def\Sym{\operatorname{Symm}}
\def\thetah{{\hat\theta}}
\def\Uh{{\widehat U}}

\def\vh{{\hat v}}

\def\xb{{\bar x}}
\def\xib{\bar \xi}
\def\xih{{\hat\xi}}

\def\Xt{{\widetilde X}}
\def\yb{{\bar y}}
\def\YO{{Y_{[0]}}}
\def\zh{{\hat z}}
\def\ZZ{{\bold Z}}

\topmatter
\title Super symplectic geometry and prequantization
\endtitle
\author G.M. Tuynman \endauthor
\rightheadtext{Super symplectic geometry and prequantization \datum}
\leftheadtext{G.M. Tuynman \datum}
\address Universit\'e de Lille I; F-59655 Villeneuve d'Ascq Cedex; France
\endaddress
\email Gijs.Tuynman\at univ-lille1.fr \endemail
\abstract 
We review the prequantization procedure in the context of super symplectic
manifolds with a symplectic form which is not necessarily homogeneous. In
developing the theory of non homogeneous symplectic forms, there is one
surprising result: the Poisson algebra no longer is the set of smooth functions
on the manifold, but a subset of functions with values in a super vector space
of dimension $1\vert1$. We show that this has no notable
consequences for results concerning coadjoint orbits, momentum maps, and
central extensions. Another surprising result is that prequantization in
terms of complex line bundles and prequantization in terms of principal circle
bundles no longer are equivalent if the symplectic form is not even.

\endabstract
\endtopmatter

\document

\heading \headnum. Introduction \endheading

Prequantization is usually seen as the first step in the geometric quantization
procedure. In \cite{Tu1} I have argued that it might be a better idea to
interpret prequantization as part of classical mechanics, at least after a
small modification. In this paper I will argue that for super
symplectic manifolds the interpretation of prequantization as part of a
quantization scheme is less obvious than one might think. In order to
understand the argument, we have to take a closer look at  the standard
prequantization procedure.

The key point in the whole argument is that there is not one single
prequantization procedure, but that there are two (equivalent) procedures. One
of them finds its origin in representation theory (the orbit method) and is
usually associated with the name of Kostant \cite{Ko1}. In this approach
prequantization of a symplectic manifold $(M,\omega)$ is an answer to the
question: find a complex line bundle $L\to M$ with a connection $\nabla$ such
that its curvature is ($i/\hbar$ times) the symplectic form
$\omega$. And as is well known, such a complex line bundle exists if and only if
$\omega/\hbar$ represents an integral cohomology class in the de~Rham cohomology
group $H^2_{dR}(M)$. The other prequantization procedure has its origin in
physics and is associated with the name of Souriau \cite{So}. For him
prequantization of a symplectic manifold is an answer to the question: find a
principal $\SS^1$-bundle $\pi:Y\to M$ with an
$\SS^1$-invariant 1-form $\alpha$ such that $\pi^*\omega = d\alpha$ and such that
$\int_{fibre} \alpha = 2\pi\hbar$. And again, such
a bundle exists if and only if $\omega/\hbar$ represents an integral cohomology
class. The equivalence with the approach by Kostant is given by taking the
associated complex line bundle relative to the canonical representation of
$\SS^1 = \{ z\in \CC \mid |z|=1\, \}$ on $\CC$. 
As Th.~Friedrich pointed out to me, one should be very careful in
translating Souriau's question in terms of connections and curvature. 
A connection 1-form on a principal fiber bundle is a Lie algebra valued 1-form
and its curvature is a Lie algebra valued 2-form. Now
it is true that the Lie algebra of $\SS^1$ is isomorphic to $\RR$, but the
isomorphism depends upon the choice of a basis vector for the Lie algebra. If
$b$ is a basis vector for the Lie algebra of $\SS^1$, Souriau's
question can be restated as: find a principal $\SS^1$-bundle $\pi:Y\to M$ with a
connection 1-form $\double\alpha$ such that $\pi^*\omega \otimes b =
d\,\double\alpha$ under the condition that the basis vector $b$ is the smallest
non-zero positively oriented vector such that $\exp(2\pi\hbar b) = 1$, where
$\exp$ denotes the exponential map from Lie algebra to Lie group. Apart from the
orientation, the condition on
$b$ is equivalent to the condition $\exp(tb) = 1
\Leftrightarrow t\in 2\pi\hbar\ZZ$. For vector bundles we do not have a similar
choice: for a vector bundle, the curvature of a connection is a 2-form with
values in the endomorphisms of the typical fiber (a vector space). And the
endomorphisms of
$\CC$ are canonically isomorphic to $\CC$.

My approach in \cite{Tu1} was to drop, in Souriau's question, the condition
$\int_{fibre} \alpha = 2\pi\hbar$ (or equivalently the condition $\exp(tb) =1
\Leftrightarrow t\in 2\pi\hbar\ZZ$ on the basis vector $b$). In this paper my
argument will be that for super symplectic manifolds Souriau's question has an
answer provided
$\omega/\hbar$ represents an integral cohomology class, whereas Kostant's
question will in general not have an answer. The argument is that the
curvature of a connection on a vector bundle must be an \stress{even} 2-form,
whereas I will consider mixed symplectic forms on supermanifolds. The reason
not to restrict attention to even symplectic forms on supermanifolds is that
coadjoint orbits of super Lie groups  will in general carry a natural mixed
symplectic form, not necessarily an even one. 

In sections 3--7 we will explain the theory of super
symplectic manifolds with mixed symplectic forms, the consequences
for the Poisson algebra, momentum maps, and central extensions, as well as the
theory of super coadjoint orbits. In sections 8--9 we
will give a detailed review of the two prequantization procedures in the
context of super symplectic manifolds.

\heading \headnum. Conventions, notation and useful results \endheading

I will work with the geometric $H^\infty$ version of DeWitt supermanifolds,
which is equivalent to the theory of graded manifolds of Leites and Kostant
(see \cite{DW}, \cite{Ko2}, \cite{Le}, \cite{Tu2}). Any reader using a
(slightly) different version of supermanifolds should be able to translate the
results to his version of supermanifolds.

\bigskip


\itemize
The basic graded ring will
be denoted as $\CA$ and we will think of it as the exterior algebra $\CA
= \Lambda V$ of an infinite dimensional real vector space $V$.

\itemize
Any element $x$ in a graded space splits into an even and an odd part $x=x_0 +
x_1$. The parity function $\e$ is defined on homogeneous elements, \ie, elements
$x$ for which either the even part $x_0$ or the odd part $x_1$ is zero. More
precisely, if $x=x_\alpha$,  then $\e(x) = \alpha$.

\itemize
On any graded space we define an involution $\invol$ by $\invol : x=x_0 + x_1
\mapsto x_0 - x_1$, where $x_\alpha$ denotes the homogeneous part of $x$ of
parity $\alpha$. If $a$ and $b$ are two elements of $\CA$ of which $a$ is
homogeneous, we then can write $a\cdot b = \invol^{\e(a)}(b) \cdot a$, meaning
that if $a$ is even, $ab = ba$ and if $a$ is odd, $ab = \invol(b) \cdot a =
(b_0 - b_1) \cdot a$. 

\itemize
All (graded) objects over the basic ring $\CA$ have an underlying real
structure, called their body, in which all nilpotent elements in $\CA$ are
ignored\slash killed. This forgetful map is called the body map, denoted by the
symbol $\body$. For the ring $\CA$ this is the  map\slash
projection $\body : \CA =\Lambda V \to \Lambda^0 V = \RR$.

\itemize
If $\omega$ is a $k$-form and $X$ a vector field, we denote the contraction of
the vector field $X$ with the $k$-form $\omega$ by $\contrf{X}{\omega}$, which
yields a $k-1$-form. If $X_1, \dots, X_\ell$ are $\ell\le k$ vector fields, we
denote the repeated contraction of $\omega$ by $\contrf{X_1, \cdots,
X_\ell}{\omega}$. More precisely:
$$
\contrf{X_1, \cdots, X_\ell}{\omega} = 
\Bigl( \contrf{X_1}{} \scirc \cdots \scirc \contrf{X_\ell}{}
\Bigr) \, \omega
\mapob.
$$
In the special case $\ell = k$ this definition differs by a factor
$(-1)^{k(k-1)/2}$ from the usual definition of the evaluation of a $k$-form on
$k$ vector fields. This difference is due to the fact that in ordinary
differential geometry repeated contraction with $k$ vector fields corresponds
to the direct evaluation in the reverse order. And indeed, $(-1)^{k(k-1)/2}$ is
the signature of the permutation changing $1,2,\dots,k$ in $k,k-1,\dots,2,1$.
However, in graded differential geometry this permutation not only introduces
this signature, but also signs depending upon the parities of the vector
fields. These additional signs are avoided by our definition.

\itemize
The evaluation of a left linear map $\mu$ on a vector $v$ is denoted as
$\contrs{v}{\mu}$. For the contraction of a multi-linear form with a vector we
will use the same notation as for the contraction of a differential form with a
vector field. In particular, we denote the evaluation of a left bilinear map
$\Omega$ on a vector $v$ by $\contrf{v}{\Omega}$, which yields a left linear map
$w\mapsto \contrs{w}{\contrf{v}{\Omega}} \equiv \contrf{w,v}{\Omega}$.

\itemize
If $E$ is \agvs/, $E^*$ will denote the \stress{left} dual of $E$, \ie, the
space of all \stress{left} linear maps from $E$ to $\CA$.

\itemize
The  general formula for evaluating the exterior derivative of a $k$-form
$\omega$ on $k+1$ vector fields $X_0, \dots, X_k$ is given by the formula
$$
\multline
(-1)^k 
\contrf{X_0,\dots,X_k}{\,d\omega} = 
\\
=   \sum_{0\le i \le k} (-1)^{i +
\underset{p<i}\to\sum 
\e({X_p})\e({X_i})} X_i(\contrf{X_0,
\dots, X_{i-1}, X_{i+1}, \dots, X_k}{\,\omega}) 
\\ 
+   \sum_{0\le i<j\le k} (-1)^{j + 
\underset{i<p<j}\to\sum 
\e({X_p})\e({X_j})}\  
\contrf{X_0, \dots,
X_{i-1},
\\
[X_i,X_j], X_{i+1}, 
\dots, X_{j-1}, X_{j+1},
\dots, X_k}{\,\omega}
\mapob
\endmultline
\tag{\memorize\theoremnummer={extderivative}}
$$
The factor $(-1)^k$ is conventional and (again) is a consequence of our
convention that $\contrf{X_0, \dots, X_k}{}$ denotes the repeated contraction
$\contrf{X_0}{} \scirc \cdots \scirc \contrf{X_k}{}$.

The special cases of a 1-form and a 2-form are sufficiently interesting to
write the definition explicitly. For a 1-form $\omega$ and homogeneous vector
fields $X$ and $Y$ we have
$$
- \contrf{X,Y}{\omega} 
=
X\bigl( \contrf{Y}{\omega} \bigr)
-
\comme XY Y\bigl( \contrf{X}{\omega} \bigr)
- \contrf{\lieb{X,Y}}{\omega}
\mapob.
\formula{extderoneform}
$$
For a 2-form $\omega$ and homogeneous vector fields $X$, $Y$,  $Z$ we have
$$
\multline
\contrf{X,Y,Z}{d\omega} 
=
X \bigl( \contrf{Y,Z}{\omega} \bigr) 
\\
- \comme XY
Y \bigl( \contrf{X,Z}{\omega} \bigr) +
\comm{\e(Z)}{(\e(X) + \e(Y))}
Z \bigl( \contrf{X,Y}{\omega} \bigr) 
\\
-
\contrf{[X,Y],Z}{\omega} +
\comme YZ 
\contrf{[X,Z],Y}{\omega} +
\contrf{X,[Y,Z]}{\omega} 
\mapob.
\endmultline
\tag\memorize\theoremnummer={extdertwoform}
$$

\itemize
The Lie derivative of a $k$-form in the direction of a vector field
$X$ is defined as usual by the formula $\Lied(X) = \contrf{X}{} \scirc{d}
+ d\scirc \contrf{X}{}$. It obeys the usual relation with the contraction
with the commutator of two vector fields $X$ and $Y$~:
$$
\contrf{\lieb{X,Y}}{} = \lieb{\Lied(X), \contrf{Y}{}} 
= \lieb{\contrf{X}{}, \Lied(Y)}
\mapob.
$$

\itemize
If $G$ is \aglgrp/, then its \glalg/ $\Liealg g$ is $\Liealg g = T_eG$, whose
Lie algebra structure is given by the commutator of left-invariant vector
fields (who are determined by their value at $e\in G$).

\itemize
If $\Phi : G \times M \to M$ denotes the (left) action of \aglgrp/ $G$ on
\agmfd/ $M$, then for all $v \in \Liealg g = T_eG$ the associated
\stress{fundamental vector field} $v^M$ on $M$ is defined as $v^M\restricted_m =
- T_{(e,m)}\Phi (v,0)$. The minus sign is conventional and ensures that the map
from $\Liealg g$ to vector fields on $M$ is a homomorphism of \glalg/s. 

Similarly, if $\Phi : M \times G \to M$ is a right action of $G$ on $M$, then
the fundamental vector field $v^M$ associated to $v\in \Liealg g$ is defined as 
$v^M\restricted_m = T_{(m,e)}\Phi (0,v)$. And again the map from $\Liealg g$ to
vector fields on $M$ is a morphism of \glalg/s. In the special case when $M =
G$ with the natural right action on itself, the fundamental vector fields are
exactly the left-invariant vector fields on $G$.


\heading \headnum. Super symplectic \gmfd/s \endheading

If we generalize naively the notion of a symplectic manifold to \gmfd/s, we
would define it as \agmfd/ $M$ with a closed and non-degenerate 2-form
$\omega$. As in the ungraded case, we would expect that the commutator of two
locally hamiltonian vector fields is globally hamiltonian. The following
example shows that this is too naive.

\example{Example}
Let $M$ be the even part of \agvs/ of dimension $2|2$ with global even
coordinates $x,y$ and global odd coordinates $\xi, \eta$. We define the closed
2-form $\omega$ by
$$
\omega = dx \wedge dy + d\xi \wedge d\eta + dx \wedge d\xi
\mapob.
$$
That $\omega$ is non-degenerate follows immediately from the equations
$\contrf{\partial_x}{\omega} = dy + d\xi$, $\contrf{\partial_y}{\omega} = -dx$, 
$\contrf{\partial_\xi}{\omega} = d\eta - dx$, and
$\contrf{\partial_\eta}{\omega} = d\xi$.

On $M$ we introduce the vector fields $X$ and $Y$ by
$$
X = 2y\partial_x - 2y\partial_\eta
\qquad\text{and}\qquad
Y = -\xi \partial_\xi + \eta\partial_\eta + \xi\partial_y
\mapob.
$$
It is immediate that $\contrf{X}{\omega} = d(y^2)$ and that $\contrf{Y}{\omega} =
d(\eta\xi)$, \ie, $X$ and $Y$ are globally hamiltonian in the naive sense. An
elementary computation shows that
$$
\lieb{X,Y} = -2\xi\partial_x -2y\partial_\eta -2\xi\partial_\eta
\mapob,
$$ 
and then it is immediate to obtain $\contrf{\lieb{X,Y}}{\omega} = d(y\xi) +
2\xi\,d\xi$, which is not closed. In other words, the commutator $\lieb{X,Y}$
of two globally hamiltonian vector fields is not even locally hamiltonian in the
naive sense.

\endexample

\definition{{Definitions}}
A 2-form $\omega$ on an \gmfd/ $M$ is called \stress{non-degenerate} if for all
$m\in M$ we have $\ker(\omega\restricted_m) = \nul$, where we interpret
$\omega\restricted_m$ as the map $X \mapsto \contrf{X}{\omega\restricted_m}$ from
$T_mM$ to $(T_mM)^*$. The 2-form $\omega$ is called \stress{homogeneously
non-degenerate} if for all $m\in M$ we have $\ker(\omega_0\restricted_m) \cap
\ker(\omega_1\restricted_m) = \nul$. Here $\omega_\alpha$ denotes the
homogeneous part of parity $\alpha$ of $\omega$, and
$\omega_\alpha\restricted_m$ is interpreted as the map $X \mapsto
\contrf{X}{\omega_\alpha\restricted_m}$ from $T_mM$ to $(T_mM)^*$. 

A 2-form $\omega$ is called \stress{symplectic} if it is closed and
homogeneously non-degenerate. A \stress{\ssmfd/} is an \gmfd/ $M$ together with
a symplectic form $\omega$.

A smooth vector field $X$ on a \ssmfd/ $(M, \omega)$ is said to be
\stress{locally\slash  globally hamiltonian} if \stress{both}
$\contrf{X}{\omega_0}$ and
$\contrf{X}{\omega_1}$ are closed\slash exact. A locally hamiltonian vector
field is sometimes called an \stress{infinitesimal symmetry}, and the set of
all locally hamiltonian vector fields is denoted as $\Sym(M,\omega)$. Its
subset of all globally hamiltonian vector fields is denoted as
$\HSym(M,\omega)$.

\enddefinition

Since $\contrf{X}{\omega} = \contrf{X}{\omega_0} + \contrf{X}{\omega_1}$, it
follows immediately that a non-degenerate 2-form is homogeneously
non-degenerate. As a consequence, a homogeneous closed 2-form $\omega$ (meaning
that either $\omega_0$ or $\omega_1$ is zero) is symplectic if and only
if it is non-degenerate, \ie, the natural definition. There are (at least) two
reasons to define a symplectic form as only being homogeneously non-degenerate.
The first is that coadjoint orbits have a natural symplectic form in this sense
which need not be non-degenerate. This is at the same time a reason to consider
non-homogeneous symplectic forms. The second reason is that it is the natural
condition that makes the definition of the Poisson algebra possible for
non-homogeneous symplectic forms. And in fact, the definition of the Poisson
algebra is slightly less straightforward in the graded case because, as we have
seen, with the naive definition of globally\slash locally hamiltonian vector
fields, the commutator of two globally hamiltonian vector fields need not even
be locally hamiltonian.

If $x^1, \dots, x^n$ are local coordinates on a \ssmfd/ $(M,\omega)$ (even and
odd together), the symplectic form can be written as $\omega = \sum_{i,j}
\omega_{ij} dx^i \wedge dx^j$ for some graded skew-symmetric matrix of local
functions $\omega_{ij}$. The classical Darboux theorem says that around every
point there exist local coordinates for which these functions $\omega_{ij}$ are
constant and of a special form. In the case of mixed symplectic forms such a
theorem is no longer possible. The simple reason is that it would imply that in
particular the rank of the even 2-form $\omega_0$ is constant. And the example
of $\omega = x\,dx \wedge dy + dx \wedge d\xi + dy \wedge d\eta$ on the \gmfd/
of dimension $2|2$ with coordinates $x,y,\xi,\eta$ shows that this need not be
the case. However, under the right circumstances we can prove an analogue of
Darboux's theorem for mixed symplectic forms. The proof is
a close copy of the Moser-Weinstein proof of Darboux's classical theorem
\cite{Wo}.

\proclaim{\thmm{MWD}{Lemma}}
Let $M$ be a \gmfd/ and let $\omega,\sigma$ be two symplectic forms on $M$. Let 
$m_o \in \body M$ a point with real coordinates such that $\omega_{m_o} =
\sigma_{m_o}$. Suppose $U$ is a neighborhood of $m_o$ with the
following properties:
\roster
\item"(i)"
there exists a 1-form $\alpha$ on $U$ such that $d\alpha = \sigma - \omega$ on
$U$ and $\alpha_{m_o} = 0$;

\item"(ii)"
there exists an open neighborhood $\Uh$ of $U \times \{s\in \CA_0 \mid 0 \le
\body s \le 1 \}$ in $U \times \CA_0$;

\item"(iii)" 
there exists an even vector field $X$ on $\Uh$
satisfying
$\contrf{X}{ds} = 1$ and $\contrf{X}{\Omega} = 0$ with $\Omega$ the closed
2-form  defined by $
\Omega_{(m,s)} = \omega_m + s \cdot (\sigma_m - \omega_m)
+ ds \wedge \alpha_m$, where $s$ is a (global even) coordinate on $\CA_0$.

\endroster
Then there exist neighborhoods $V,W \subset U$ of $m_o$ and a diffeomorphism
$\rho : V \to W$ such that $\rho^*\sigma =\omega$.

\endproclaim

\dontprint{%
\demo{Proof}
Without loss of generality we may assume that $U$ is a coordinate chart with
coordinates $x^1, \dots, x^n$. The condition $\contrf{X}{ds} = 1$ implies that
$X$ is of the form $X_{(x,s)} = \partial_s + Y_{(x,s)}$ with $Y_{(x,s)} =\sum_i
X^i(x,s) \partial_{x^i} $. 
Using that $\omega_{m_o} = \sigma_{m_o}$ and that $\alpha_{m_o} = 0$, the
condition $\contrf{X}{\Omega} = 0$ gives us $\contrf{Y_{(m_o,s)}}{\omega_{m_o}}
= 0$. We denote by $\phi_t$ the flow of the even vector field $X$ (see
\cite{Tu2} for more details on integrating even vector fields on
\gmfd/s).
Since the coefficient of $\partial_s$ is 1, $\phi_t$ is necessarily of the form
$\phi_t(m,s) = (??,s+t)$. 
Since $Y$ is even and $\omega$ homogeneously non-degenerate, it follows
that $Y_{(m_o,s)} = 0$ for all $s$. It follows that the integral curve of $X$
through $(m_o,0)$ is defined at least for all $t\in \CA_0$ such that $0\le
\body t \le 1$ (because $\phi_t(m_o,0) = (m_o,t)$,
which remains in $\Uh$ for these values of $t$). Since the domain of the flow
is open and the interval $[0,1]$ compact, there exists a neighborhood $V\subset
U$ of $m_o$ and a neighborhood $I$ of $0\in \CA_0$ such that $\phi_t(m,s)$ is
defined for all $(m,s) \in V \times I$ and all $0\le \body t \le 1$. We now
define $\rho : V \to U$ by the equation $\phi_1(m,0) = (\rho(m) , 1)$. This is a
diffeomorphism onto $W = \rho(V)$ with inverse given by the equation
$\phi_{-1}(m,1) = (\rho^{-1}(m) ,0)$. 

Since $\Omega$ is closed and $\contrf{X}{\Omega} = 0$, we have $\Lied(X)\Omega =
0$, and thus the flow $\phi_t$ preserves $\Omega$~: $\phi_t^*\Omega = \Omega$.
We now denote by $i_0, i_1:U \to \Uh$ the canonical injections $i_j(m) =
(m,j)$, $j=0,1$. By definition we have $\phi_1 \scirc i_0 = i_1 \scirc \rho$,
but also $i_1^*\Omega = \sigma$ and $i_0^*\Omega = \omega$. We then compute: 
$$
\rho^*\sigma 
= 
(i_1 \scirc \rho)^*\Omega
=
(\phi_1 \scirc i_0)^*\Omega
=
i_0^*(\phi_1^*\Omega) 
=
i_0^*\Omega = \omega
\mapob.
\eqno\QEDbox
$$
\enddemo
} 

In general it will not be easy to satisfy the conditions of \recalt{MWD}.
However, in the special case that $\omega$ is homogeneous, the conditions can be
fulfilled (see also \cite{Ko2}).

\proclaim{\thm{Proposition}}
Let $\omega$ be a homogeneous symplectic form on a connected \gmfd/ $M$ of
dimension
$p|q$ and let $m_o\in \body M$ be arbitrary.

If $\omega$ is even, then there exist $k,\ell \in \NN$, $p=2k$ (\ie, $p$ is
even), $0\le \ell\le q$ and a coordinate neighborhood $U$ of
$m_o$ with coordinates
$x^1, \dots, x^k, y_1, \dots,y_k, \xi^1, \dots, \xi^q$ ($x,y$ even and $\xi$
odd) such that
$\omega = \sum_{i=1}^k dx^i \wedge dy_i + \sum_{i=1}^\ell d\xi^i \wedge d\xi^i
- \sum_{i=\ell+1}^q d\xi^i \wedge d\xi^i$ on $U$.

If $\omega$ is odd, then $p=q$ and there exists a coordinate neighborhood $U$ of
$m_o$ with coordinates $x^1, \dots, x^p, \xi^1, \dots, \xi^p$ ($x$ even and
$\xi$ odd) such that $\omega = \sum_{i=1}^p dx^i \wedge d\xi^i$ on $U$.

\endproclaim

\dontprint{
\demo{Proof}
Let $x^1, \dots, x^p,
\allowbreak\xi^1, \dots, \xi^q$ be a coordinate system on a chart
$O$ around $m_o$. 

\itemize
{\it The even case.\/}
At $m_o$ $\omega$ has the form $\omega_{m_o} = 
{\sum_{ij} A_{ij} dx^i \wedge dx^j} + \sum_{ij} S_{ij} d\xi^i \wedge d\xi^j$ for
some real matrices $A$ and $S$, $A$ skew-symmetric and $S$ symmetric
($\omega$ is even and $m_o$ has real coordinates, hence mixed terms $dx^i
\wedge d\xi^j$ have zero coefficients). Since $\omega$ is homogeneous, it is
non-degenerate, hence both $A$ and $S$ are invertible. By a linear change of
coordinates we thus may assume that $A$ is the canonical symplectic matrix of
size $2k \times 2k$ (proving that $p=2k$ is even) and that $S$ is the diagonal
matrix $diag(1, \dots,1, -1, \dots, -1)$ with $\ell$ plus signs and $q-\ell$
minus signs, $\ell$ being the signature of the metric $S$. Renaming the
coordinates $x^{k+1}, \dots, x^p$ to $y_1, \dots, y_k$ we thus can define the
symplectic form $\sigma$ on $O$ by ${\sigma = \sum_{i=1}^k dx^i \wedge dy_i +
\sum_{i=1}^\ell d\xi^i \wedge d\xi^i - \sum_{i=\ell+1}^q d\xi^i \wedge d\xi^i}$.
We then have by construction
$\omega_{m_o} = \sigma_{m_o}$.

Since both $\omega$ and $\sigma$ are closed, we may assume (by taking a smaller
$O$ if necessary) that there exists an even 1-form $\alpha$ on $O$ such that
$\sigma - \omega = d\alpha$ on $O$. By changing $\alpha$ by the gradient of a
function we also may assume that $\alpha_{m_o} = 0$. 

In order to find $U$, $\Uh$ and $X$ satisfying conditions (ii) and (iii) of
\recalt{MWD}, we first note that on $O \times \CA_0$ the condition
$\contrf{X}{ds} = 1$ implies that $X$ is of the form $X_{(x,s)}  = \partial_s +
Y_{(m,s)}$ with $Y_{(m,s)} =\sum_{i=1}^{p+q} X^i(m,s) \partial_{z^i} $ where
the $z^i$ denote all (even and odd) coordinates on $O$. The condition
$\contrf{X}{\Omega} = 0$ translates into the two equations
$\contrf{Y}{\bigl(\omega + s(\sigma-\omega) \bigr)} = -
\alpha$ and $\contrf{Y}{\alpha} = 0$. However, since $Y$ is supposed to be
even,
$\contrf{Y,Y}{\sigma} = \contrf{Y,Y}{\omega} = 0$ by skew-symmetry of 2-forms.
Hence the second condition $\contrf{Y}{\alpha} = 0$ is a consequence of the
first. We now introduce the linear maps $A(m,s) : T_mM \to (T_mM)^*$ by
$\contrs{Y}{A(m,s)} = \contrf{Y}{\bigl(\omega_m + s(\sigma_m-\omega_m)
\bigr)}$. Since $\omega$ and $\sigma$ are even, $A(m,s)$ is even. Since
$A(m_o,s) = \omega_{m_o}$ is invertible ($\omega$ is non-degenerate) for all
$s$, there exist neighborhoods $W_s$ of $m_o$ and $I_s$ of $s$ such that
$A(m,t)$ is invertible for all $(m,t) \in U_s \times I_s$. Hence, by
compactness of $[0,1]$, there exists a neighborhood $U$ of $m_o$ and a
neighborhood $I$ of $\{s\in\CA_0 \mid 0\le \body s\le1\}$ such that $A(m,t)$ is
invertible for all $(m,t) \in U \times I$. Taking $\Uh = U \times I$ and
$Y_{(m,s)} = - \contrs{\alpha_m}{A(m,s)^{-1}}$ then satisfies the conditions
because $\alpha$ and $A(m,s)$ are even and thus this $Y$ is too. Hence by
\recalt{MWD} there exist neighborhoods $V$ and $W$ of $m_o$ and a
diffeomorphism $\rho:V \to W$ such that $\rho^*\sigma = \omega$. Composing the
coordinates $z^i$ on $O$ with $\rho$ gives us the desired coordinate system.

\itemize 
{\it The odd case.\/}
At $m_o$ $\omega$ has the form $\omega_{m_o} = 
\sum_{ij} A_{ij} dx^i \wedge d\xi^j$ for
a real matrix $A$ ($\omega$ is odd and $m_o$ has real coordinates, hence the
terms $dx^i \wedge dx^j$ and $d\xi^i 
\wedge d\xi^j$ have zero coefficients). Since $\omega$ is homogeneous, it is
non-degenerate, hence $A$ must be a square invertible matrix. In particular
$p=q$. By a linear change of coordinates we thus may assume that $A$ is the
identity. We thus define the symplectic form $\sigma$ on $O$ by ${\sigma =
\sum_{i=1}^p dx^i \wedge d\xi^i}$.
We then have by construction
$\omega_{m_o} = \sigma_{m_o}$.

Since both $\omega$ and $\sigma$ are closed and odd, we may assume that there
exists an odd 1-form $\alpha$ on $O$ such that
$\sigma - \omega = d\alpha$ on $O$ (no need to take a smaller $O$ because odd
closed forms are always exact). By changing
$\alpha$ by the gradient of a function we also may assume that $\alpha_{m_o} =
0$.  In order to find $U$, $\Uh$ and $X$ satisfying conditions (ii) and (iii) of
\recalt{MWD}, we proceed exactly as in the even case. The only difference is
that here $\alpha$ and $A(m,s)$ are odd. But then again $Y$ is even.
\QED\enddemo
} 

We have defined a symplectic form as a homogeneously non-degenerate closed
2-form. For the sequel it is important to note that a different interpretation
is possible. The trick we will use is quite general: it is a way to transform
any non-homogeneous graded object in an even object. We fix once and for all 
\agvs/ $C$ of dimension $1|1$ with basis $c_0, c_1$ with (of course) parities
$\e(c_\alpha) = \alpha$. For any $k$-form $\omega = \omega_0 + \omega_1$ on an
\gmfd/ $M$ we then can define the \stress{even} $C$-valued $k$-form
$\double\omega$ by $\double\omega = \omega_0 \otimes c_0 + \omega_1 \otimes c_1$. The map
$\omega \mapsto \double\omega$ establishes a bijection between $k$-forms $\omega$ and
even $C$-valued $k$-forms $\double\omega$. We will define the exterior derivative of
a $C$-valued $k$-form component wise, as well as the notions of  closed\slash
exact $C$-valued $k$-forms. In particular, a $C$-valued $1$-form $\Omega$ is
exact if and only if there exists a (smooth) function $F:M\to C$ (\ie, a
$C$-valued $0$-form) such that $\Omega = dF$.

\definition{Alternative definitions}
A closed 2-form
$\omega$ is symplectic if and only if for all $m\in M$ we have $\ker
\double\omega\restricted_m = \nul$, where we interpret $\double\omega\restricted_m$ as the
map $X \mapsto \contrf{X}{\double\omega\restricted_m}$ from $T_mM$ to $(T_mM)^* \otimes
C$. Moreover, a vector field $X$ on a symplectic manifold $(M,\omega)$ is
locally\slash globally hamiltonian if and only if $\contrf{X}{\double\omega}$ is
closed\slash exact.

\enddefinition

\remark{Remark}
The construction of $\double\omega$ is not intrinsic but depends upon the choice of
the basis $c_0, c_1$ for $C$. One can make it more intrinsic by starting from
\agvs/ $E$ of dimension $1|0$ and to define $C$ as $C = E \oplus E^\flat$,
where $E^\flat \equiv \prod E$ denotes $E$ with all its parities reversed. In
this way $C$ depends only upon the choice of a single basis vector for $E$.
We can even get rid of this last arbitrariness by choosing $E = \CA$, which has
a canonical basis vector $1\in \CA$. However, since later on we will interpret
$C$ as \aglalg/ and $c_0,c_1$ as a particular basis of this \glalg/, we will
keep for the moment the arbitrariness in our construction.

\endremark

\proclaim{\thmm{lochamprop}{Lemma}}
Let $(M,\omega)$ be a \ssmfd/ and  $X$  a vector field on $M$.
\roster
\item"(i)"
$X$ is locally hamiltonian 
$\Longleftrightarrow$
$\forall\alpha:\Lied(X)\omega_\alpha = 0$ 
$\Longleftrightarrow$
$\forall\alpha,\beta:\Lied(X_\beta)\omega_\alpha = 0$ 
$\Longleftrightarrow$
$\forall\alpha:\contrf{X_\alpha}{\omega}$ is closed
$\Longleftrightarrow$
$\forall\alpha:\Lied(X_\alpha)\omega = 0$
$\Longleftrightarrow$
$\Lied(X)\double\omega = 0$
$\Longleftrightarrow$
$\forall\beta : \Lied(X_\beta)\double\omega = 0$,
and each of these conditions implies
$\Lied(X)\omega = 0 = d\contrf{X}{\omega}$.

\item"(ii)"
$X$ is globally hamiltonian
$\Longleftrightarrow$
$\forall \alpha,\beta : \contrf{X_\alpha}{\omega_\beta}$ is exact
$\Longleftrightarrow$
$\forall\alpha : \contrf{X_\alpha}{\omega}$ is exact
$\Longleftrightarrow$
$\forall\alpha : \contrf{X_\alpha}{\double\omega}$ is exact,
and each of these conditions implies that
$X$ is locally hamiltonian and that $\contrf{X}{\omega}$ is exact.

\item"(iii)"
Given a smooth function $f:M\to C$, there exists at most one
vector field $X$ on $M$ such that
$\contrf{X}{\double\omega} = df$.

\endroster

\endproclaim

\dontprint{
\demo{Proof}
For parts (i) and (ii), we note that $\Lied(X) = \lieb{d\,,\contrf{X}{}} = d
\scirc \contrf{X}{} + \contrf{X}{} \scirc d$ and $d\omega_\alpha = 0$. From this
the first part follows immediately. Since the homogeneous parts of a closed
1-form are separately closed, it follows that the homogeneous parts of
$\contrf{X}{\omega_\beta}$, which are  $\contrf{X_\alpha}{\omega_\beta}$, are
closed. And then we can sum over $\beta$ to obtain
$d\contrf{X_\alpha}{\omega} = 0$. The other implications of (i) and (ii) follow
similarly. 

For part (iii) let $X$ and $Y$ be two such vector fields, then we have
$\contrf{X-Y}{\double\omega} = 0$. Since $\ker\double\omega = \nul$, it follows that $X-Y =
0$.
\QED\enddemo
} 

\remark{Remark}
The usefulness of the condition that a symplectic form be homogeneously
non-degenerate shows itself in property (iii) of \recalt{lochamprop}: it is the
natural condition to guarantee uniqueness of $X$.

\endremark

\definition{Definitions}
The \stress{Poisson algebra} $\Poisson$ of a \ssmfd/ $(M, \omega)$ is defined as
a subset of $\Ci(M,C)$ by
$$
\Poisson = \{ f\in \Ci(M,C) \mid
\exists X 
 : \contrf{X}{\double\omega} = df  \,\}
\mapob.
$$
For $f\in \Ci(M,C)$ we denote by $f^\alpha\in \Ci(M)$ the components of $f$ with
respect to the basis $c_0,c_1$ of $C$, \ie, $f = f^0 c_0 + f^1 c_1$. On the
other hand, the homogeneous parts of $f$ are denoted as usual by $f_\alpha \in
\Ci(M,C)$, \ie, $\forall m\in M : f_\alpha(m) \in C_\alpha$. If we decompose
each $f_\alpha$ according to the basis $c_0, c_1$, we get four homogeneous
functions $f^\alpha_\beta \in \Ci(M)$ with 
$$
f_0 = f^0_0c_0 +  f^1_0 c_1
\qquad\text{and}\qquad
f_1 = f^0_1 c_0 + f^1_1 c_1
\mapob.
$$
Taking the parities of $c_\alpha$ into account, we have $\e(f^\alpha_\beta) =
\alpha + \beta$. In particular, if we decompose the coefficient functions
$f^\alpha$ into their homogeneous parts, we get
$$
(f^\alpha)_\beta = f^\alpha_{\alpha+\beta}
\mapob.
$$
The Poisson algebra
$\Poisson$ becomes \arvs/ when we define addition and
multiplication by reals in the natural way.

According to \recalt{lochamprop} there can only be one $X$ satisfying the
conditions for $f \in \Ci(M,C)$ to belong to
$\Poisson$. This unique vector field is denoted as $X_f$ and is
called the \stress{hamiltonian vector field associated to $f \in
\Poisson$.} Splitting the defining equation for the
hamiltonian vector field $X \equiv X_f$ in the homogeneous parts of
the components gives us the equations $\contrf{X_\beta}{\omega_\alpha} =
df^\alpha_{\beta}$, simply because the parity of
$\contrf{X_\beta}{\omega_\alpha}$ is
$\alpha + \beta$, as is the parity of $f^\alpha_{\beta}$.

The \stress{Poisson bracket} $\PB{\,f\,,\,g\,}$ of two
elements $f$, $g \in \Poisson$ is defined as 
$$
\PB{\,f\,,\,g\,}
= \contrf{X_f,X_g}{\double\omega} \equiv
\contrf{X_f}{\bigl( \contrf{X_g}{\double\omega} \bigr)} = X_f g
\mapob,
$$
where for the last equality we defined the action of a vector field on a
$C$-valued function component wise and where we used the defining equation
$\contrf{X_g}{\double\omega} = dg$.

\enddefinition

\remark{Remark}
If the symplectic form $\omega$ is homogeneous and if $M$ is connected, then the
Poisson algebra $\Poisson$ is isomorphic to $\Ci(M) \times \RR$. For instance,
let $\omega$ be even, \ie, $\omega_1 = 0$ and $\omega_0 = \omega$ is
non-degenerate in the usual sense. It follows that $f^0c_0 + f^1c_1$ belongs to
$\Poisson$ if and only if there exists a vector field $X$ on $M$ such that
$\contrf{X}{\omega_0} = df^0$ and $df^1 = 0$. Non-degeneracy of $\omega_0$
implies that such an $X$ exists for all $f^0$ and connectedness of $M$ implies
that $f^1$ must be constant.

From the above analysis, the reader (just as the author) might get the idea
that the map $\Poisson \to \Ci(M)$, $f\mapsto f^0 + f^1$ only has constant
functions in its kernel (and that it might be surjective). However, the
following example (a coadjoint orbit!) shows that such a belief is false. 

\endremark

\example{Example}
Let 
$M$ be the even part of
\agvs/ of dimension $2|1$ with even coordinates
$x,y$, and odd coordinate $\xi$. Then the closed 2-form $\omega = 
{dx \wedge dy} + dx \wedge d\xi$ is degenerate but homogeneously non-degenerate,
\ie,  symplectic. 
For $f = {f^0c_0 + f^1c_1} \in \Ci(M,C)$ and $X = X^x \partial_x + X^y
\partial_y + X^\xi \partial_xi$, the condition that $f$ belongs to $\Poisson$
translates as the conditions
$$
X^x dy - X^y dx = df^0
\quad,\quad
X^x d\xi - X^\xi dx = df^1
\mapob.
$$
From this it follows that $f^0$ is independent of $\xi$, that $f^1$ is
independent of $y$ and that we have
$$
X^x = \fracp{f^0}y = \fracp{f^1}{\xi}
\quad,\quad
X^y = -\fracp{f^0}x
\quad,\quad
X^\xi = -\fracp{f^1}x
\mapob.
$$
Since $f^1$ is independent of $y$, we can write $f^1(x,y,\xi) = f^1_0(x) + \xi
f^1_1(x)$ for smooth functions $f^1_0, f^1_1$ of $x$. But then
$\partial_y{f^0}(x,y) = f^1_1(x)$ is independent of $y$, and thus $f^0(x,y) =
f^0_0(x) + yf^1_1(x)$ for some function $f^0_0$ of $x$. We conclude that $f\in
\Poisson$ is of the form
$$
f(x,y,\xi) = \bigl( f^0_0(x) + yf^1_1(x) \bigr) \cdot c_0 + \bigl( f^1_0(x) + \xi
f^1_1(x) \bigr) \cdot c_1
\mapob.
$$
In other words, $\Poisson \cong [\Ci(\CA_0)]^3$. It follows that the kernel of
the map $\Poisson \to \Ci(M)$, $f\mapsto f^0+f^1$ is isomorphic to $\Ci(\CA_0)$
and that its image is isomorphic to $[\Ci(\CA_0)]^2$. 

\endexample

\proclaim{\thmm{commlocisglob}{Lemma}}
On a \ssmfd/ $(M,\omega)$ the commutator $[X,Y]$ of two locally hamiltonian
vector fields $X$ and $Y$ is the (globally) hamiltonian vector field associated
to $\contrf{X,Y}{\double\omega} \in \Poisson$.

\endproclaim

\dontprint{
\demo{Proof}
Using \recalt{lochamprop} we compute:
$$
\eqalignno{
\contrf{[X,Y]}{\double\omega} &=
[\,\Lied(X)\,,\,\contrf{Y} \,] \, \double\omega
\cr &=
\Lied(X) \contrf{Y}{\double\omega} - 
\sum_{\alpha,\beta = 0}^1 
\comm\alpha\beta
\contrf{Y_\beta}{\Lied(X_\alpha)\double\omega}
\cr &=
d\contrf{X}{\contrf{Y}{\double\omega}}
+ \contrf{X}{d\contrf{Y}{\double\omega}}
=d\contrf{X}{\contrf{Y}{\double\omega}}
\mapob.&\QEDbox
} 
$$
\enddemo
} 

\proclaim{\thm{Lemma}}
Let $\Poisson$ be the Poisson algebra of a \ssmfd/ $(M,\omega)$.
The Poisson bracket on $\Poisson$ is a well defined even bracket which gives
$\Poisson$ the structure of \arglalg/. Moreover, the map
$f \mapsto X_f$ from $\Poisson$ to hamiltonian vector fields is an
even morphism of \rglalg/s. Explicitly:
\roster
\item"(i)" For $f,g \in \Poisson$ we have  $[\,X_f \,,\,
X_g
\,] = X_{\PB{\,f\,,\,g\,}}$.

\item"(ii)"
The bracket is bilinear;

\item"(iii)"
For homogeneous $f,g \in \Poisson$ we have
$\PB{\,f\,,\,g\,} = - \comme fg
\PB{\,g\,,\,f\,}\,$;

\item"(iv)"
For homogeneous $f,g,h \in \Poisson$ we have 
$$
\multline
\comme fh \PB{\,f\,,\,\PB{\,g\,,\,h\,}\,} 
+\comme gf \PB{\,g\,,\,\PB{\,h\,,\,f\,}\,} 
\\
+\comme hg \PB{\,h\,,\,\PB{\,f\,,\,g\,}\,} =0
\mapob.
\endmultline
$$

\endroster

\endproclaim

\dontprint{
\demo{Proof}
We start by proving that the map $f \mapsto X_f$ is even.
We have $\contrf{X_f}{\double\omega} = df$. If $f$ is even, non-degeneracy of
$\double\omega$ (and the fact that it is even) implies that the odd part of $X_f$ must
be zero. Since  the odd case is similar, this proves that the map $f
\mapsto X_f$ is even. Property (i) then is an immediate consequence of
\recalt{commlocisglob}. 

Since the map $f \mapsto X_f$ is even, it is immediate that the bracket on
$\Poisson$ is even, \ie, if $f,g\in \Poisson$ are homogeneous, then $\PB{f,g}$
is homogeneous of parity $\e(f) + \e(g)$. Properties (ii) and (iii) follow
immediately from the defining equations of the bracket and the fact that a 2-form
is graded skew-symmetric. 

To prove property (iv) we first establish some useful identities. We start
with the fact that by definition of the bracket and by property (i) we have
$$
\PB{f,\PB{g,h}} = \contrf{X_f, \lieb{X_g,X_h}\,}{\double\omega}
\mapob.
\formula{triplepoissonbracket}
$$
Since $d\contrf{X_f, X_g}{\double\omega} =
\contrf{\lieb{X_f,X_g}\,}{\double\omega}$ by \recalt{commlocisglob}, we have for
any vector field $Z$ on $M$ the equality
$$
Z\bigl( 
\contrf{X_f,X_g}{\double\omega}
\bigr)
=
\contrf{Z,\lieb{X_g,X_h}\,}{\double\omega}
\mapob.
\formula{vfondcontr}
$$
If we apply \recalf{vfondcontr} and \recalf{extdertwoform} to the closed
$C$-valued 2-form $\double\omega$ and the three homogeneous vector fields $X_f$,
$X_g$, $X_h$, and if  we use the graded skew-symmetry of a 2-form and of the
commutator bracket, we obtain
$$
\align
0&= 
\tfrac12
\comme fh
\contrf{X_f, X_g, X_h}{d\double\omega}
\\
&=
\comme fh
\contrf{X_f, \lieb{X_g, X_h}}{\double\omega}
+ \comme gf
\contrf{X_g, \lieb{X_h, X_f}}{\double\omega}
\\
&\qquad\qquad +
\comme hg
\contrf{X_h, \lieb{X_f, X_g}}{\double\omega}
\mapob.
\endalign
$$
Comparing this with \recalf{triplepoissonbracket} gives us property (iv).
\QED\enddemo
}

\heading \headnum. Central extensions: general theory \endheading

\definition{\thm{Definition}}
Let $\Liealg g$ be \aglalg/ and let $E$ be \agvs/. A \stress{$k$-chain on
$\Liealg g$ with values in (the trivial $\Liealg g$-module) $E$} is an even left
$k$-linear graded skew-symmetric map on $\Liealg g$ with values in $E$, \ie,
an even linear map $\bigwedge^k \Liealg g \to E$. The set of 
all such $k$-chains is denoted by $C^k(\Liealg g, E)$. On the set of $k$-chains
we define a coboundary operator $d: C^k(\Liealg g, E) \to C^{k+1}(\Liealg g, E)$
by the formula
$$
\multline
\contrf{v_0, \dots, v_k}{\,dc} = 
\\
 (-1)^k \cdot \sum_{0\le i<j\le k} (-1)^{j + 
\underset{i<p<j}\to\sum 
\e({X_p})\e({X_j})}\  
\contrf{v_0, \dots, v_{i-1},
\\
[v_i,v_j], v_{i+1}, \dots 
, v_{j-1}, v_{j+1},
\dots, v_k}{\,c}
\mapob.
\endmultline
$$

\enddefinition

\remark{\thm{Remark}}
The global factor $(-1)^k$ in the definition of the coboundary operator is
conventional and comes in because we interpret the contraction $\contrf{v_0,
\dots, v_k}{}$ as repeated contractions with a single vector: $\contrf{v_0,
\dots, v_k}{} = \contrf{v_0}{} \scirc \cdots \scirc \contrf{v_k}{}$, just as in
the case of differential forms.

\endremark

\proclaim{\thm{Lemma}}
If $G$ is an \glgrp/ whose associated \glalg/ is $T_eG = \Liealg g$, then
$C^k(\Liealg g, E)$ can be identified with the set of all even left-invariant
$k$-forms on $G$ with values in $E$, the identification given by taking the
value at $e\in G$. In this identification, the coboundary operator corresponds
to the exterior derivative.

\endproclaim

\dontprint{
\demo{Proof}
A $k$-form with values in $E$ is for each $g\in G$ a left $k$-linear graded
skew-symmetric map on $T_gG$ with values in $E$, and thus in particular at
$e\in G$ it is a $k$-chain. Conversely, any $k$-chain, \ie, a  left $k$-linear
graded skew-symmetric map on $T_eG$ with values in $E$, determines by left
translations a left-invariant $k$-form on $G$ with values in $E$.

The relation with the exterior derivative follows immediately when we apply the
general formula \recalf{extderivative} for evaluating the exterior derivative of
a $k$-form $\omega$ on $k+1$ vector fields
to the case of a left invariant $k$-form $c$ with values in $E$ evaluated on
$k+1$ left-invariant vector fields (in which case all terms in the single
summation yield zero because vector fields applied to a constant function gives
zero).
\QED\enddemo
} 

\proclaim{\thm{Corollary}}
$d^2 = d \scirc d : C^k(\Liealg g
, E) \to C^{k+2}(\Liealg g , E)$ is the zero map: $d^2 = 0$.

\endproclaim

\definition{\thm{Definition}}
Associated to the set of $k$-chains $C^k(\Liealg g, E)$ we define the set 
$Z^k(\Liealg g, E) \subset C^k(\Liealg g, E)$ as 
$$
Z^k(\Liealg g, E)  = \ker(d:C^k
\to C^{k+1})
\mapob;
$$
its elements are called \stress{$k$-cocycles}. We also define
the set $B^k(\Liealg g, E) \subset Z^k(\Liealg g, E)$ as 
$$
B^k(\Liealg g, E) =
\im(d:C^{k-1} \to C^k)
$$ 
(with $B^0(\Liealg g, E) = \nul$), whose elements are
called \stress{$k$-coboundaries}.  The quotient 
$$
H^k(\Liealg g, E) =
Z^k(\Liealg g, E)/B^k(\Liealg g, E)
$$ 
is called the \stress{Lie algebra
cohomology in dimension $k$ of $\Liealg g$ with values in (the trivial $\Liealg
g$-module) $E$}.

\enddefinition

\definition{\thm{Definition}}
Let $\Liealg g$ be an \glalg/ and let $E$ be \agvs/, which we interpret as an
abelian \glalg/ by taking the trivial (zero) bracket. 
\stress{A central extension of $\Liealg g$ by
$E$} is an exact sequence $\nul \to E \to \Liealg h \to \Liealg g \to \nul$ of
\glalg/ morphisms such that the image of $E$ lies in the center of
$\Liealg h$. If the morphisms are understood, one also calls $\Liealg h$ the
central extension of $\Liealg g$ by $E$. Two central extensions $\Liealg h$ and
$\Liealghh$ are called \stress{equivalent} if there exists an isomorphism of
\glalg/s $\phi : \Liealg h \to \Liealghh$ such that the following diagram is
commutative:
$$
\matrix
&&&&\Liealg h \\
&&&\nearrow&&\searrow \\
\nul & \to & E& 
&\updownarrow &&\Liealg g & \to & \nul \mapob.
\\ &&&\searrow && \nearrow  \\
&&&& \Liealghh
\endmatrix
\formula{equivofcentrext}
$$

\enddefinition

\example{Example}
Let $(M,\omega)$ be a connected symplectic manifold with its Poisson algebra
$\Poisson$. Obviously the constant functions belong to $\Poisson$, \ie, $\body
C \equiv \RR c_0 + \RR c_1 \subset \Poisson$. Apart from being infinite
dimensional, the exact sequence
$$
\nul \to \body C \to \Poisson @>f\,\mapsto\, X_f>>
\HSym(M,\omega)
\to \nul
$$
is a central extension of $\HSym(M,\omega)$ by $\body C$ (in the category of
\rglalg/s). That $\body C$ is the kernel of the map $f\mapsto X_f$ is a
consequence of the connectedness of $M$.

\endexample

\definition{\thmm{centralextbycocycle}{Construction}}
For any 2-cocycle $\Omega \in Z^2(\Liealg g, E)$ we define a central extension
$\Liealg h$ of $\Liealg g$ by $E$ as follows. As \gvs/ we define $\Liealg h =
\Liealg g \times E$, with maps $i: E \to \Liealg h$ and $\pi : \Liealg h \to
\Liealg g$ defined by $i(e) = (0,e)$ and $\pi(v,e) = v$. This gives us an exact
sequence $\nul \to E \to \Liealg h \to \Liealg g \to \nul$ of \gvs/s. On
$\Liealg h$ we define the bracket by
$$
\lieb{(v,e), (w,f)} = (\,\lieb{v,w}, \contrf{v,w}{\Omega}\,)
\mapob.
$$

\enddefinition

\proclaim{\thm{Lemma}}
The exact sequence defined in \recalt{centralextbycocycle} is a well defined
central extension of $\Liealg g$ by $E$. Moreover, this construction induces an
isomorphism between $H^2(\Liealg g, E)$ and equivalence classes of central
extensions of $\Liealg g$ by $E$.

\endproclaim

\dontprint{
\demo{Proof}
Since $E$ is considered to be an abelian \glalg/, it is immediate that $i:E \to
\Liealg h$ is a morphism of \glalg/s. It is also immediate from the definition
that $\pi : \Liealg h \to \Liealg g$ is a morphism of \glalg/s with $E$ as
kernel. It thus remains to show that the bracket is indeed a well defined bracket
of 
\aglalg/. By construction it is even (because $\Omega$ is even), bilinear and
graded skew symmetric. The graded Jacobi identity for three homogeneous elements
$(u,e), (v,f), (w, g) \in \Liealg h$ translates to the equation
$$
\comme uw \contrf{u, \lieb{v,w}}{\Omega} +
\comme vu \contrf{v, \lieb{w,u}}{\Omega} +
\comme wv \contrf{w, \lieb{u,v}}{\Omega} = 0
\mapob.
$$
Using the graded skew-symmetry of $\Omega$ and of the bracket on $\Liealg g$
this can be rewritten as
$$
-  \contrf{\lieb{u,v}, w}{\Omega}
+
 \comme vw \contrf{\lieb{u,w}, v}{\Omega}
+
 \contrf{u, \lieb{v,w}}{\Omega} = 0
\mapob,
$$
which is exactly the condition $d\Omega = 0$ for $\Omega$ to be a 2-cocycle.

To prove that the map $\Omega \mapsto \Liealg h$ induces an isomorphism between 
$H^2(\Liealg g, E)$ and equivalence classes of central
extensions of $\Liealg g$ by $E$, we proceed as follows. First suppose that
$\Omega'$ and $\Omega$ determine the same cohomology class, \ie, $\Omega' =
\Omega + dF$ for some $F\in C^1(\Liealg g, E)$, \ie, $F$ is an even linear
map $\Liealg g\to E$. We then define $\phi: \Liealg h = \Liealg g \times E \to
\Liealg h' = \Liealg g \times E$ by $\phi(v,e) = (v,e+\contrf{v}{F})$ and we
claim that this $\phi$ makes the two central extensions $\Liealg h$ and $\Liealg
h'$ equivalent. Obviously this $\phi$ is an isomorphism of \gvs/s and makes
\recalf{equivofcentrext} commutative. It thus remains to show that it is a
morphism of \glalg/s. We thus compute:
$$
\align
\phi(\lieb{(v,e), (w,f)}_{\Liealg h})
&=
\phi((\lieb{v,w}, \contrf{v,w}{\Omega}))
=
(\lieb{v,w}, \contrf{v,w}{\Omega} + \contrf{\lieb{v,w}}{F} )
\\
&=
(\lieb{v,w}, \contrf{v,w}{\Omega} + \contrf{{v,w}}{dF} )
=
\lieb{(v,e),(w,f) }_{\Liealg h'}
\mapob.
\endalign
$$
It follows that we indeed have a map defined on $H^2(\Liealg g, E)$.

Next suppose that $\Omega, \Omega' \in Z^2(\Liealg g, E)$ define equivalent
central extensions. We thus have an isomorphism $\phi: \Liealg h \to \Liealg
h'$ making the diagram \recalf{equivofcentrext} commutative. Commutativity
implies immediately that $\phi$ must be of the form $\phi(v,e) = (v,e +
\contrf{v}{F})$ for some even linear map $F: \Liealg g \to E$. Using the fact
that $\phi$ is a morphism of \glalg/s, we compute:
$$
\align
(\lieb{v,w}, \contrf{v,w}{\Omega'})
&=
\lieb{(v,\contrf{v}{F}), (w,\contrf{w}{F}) }_{\Liealg h'}
=
\lieb{\phi(v,0), \phi(w,0)   }_{\Liealg h'}
\\&
=
\phi(\lieb{(v,0), (w,0)   }_{\Liealg h})
=
\phi(\lieb{v,w}, \contrf{v,w}{\Omega})
\\&
=
(\lieb{v,w}, \contrf{v,w}{\Omega} + \contrf{\lieb{v,w}}{F} )
=
(\lieb{v,w}, \contrf{v,w}{(\Omega+dF)} )
\mapob.
\endalign
$$
We conclude that for all $v,w\in \Liealg g$ we have $\contrf{v,w}{(\Omega +
dF)} 
 = \contrf{v,w}{\Omega'}$, \ie, $\Omega +dF = \Omega'$. In other words, if
$\Omega$ and $\Omega'$ determine equivalent central extensions, they are
cohomologous, and thus the map from $H^2(\Liealg g, E)$ to equivalence classes of
central extensions of $\Liealg g$ by $E$ is injective.

To prove surjectivity, let $\nul \to E 
\overset{i}\to{\to} \Liealg h 
\overset{\pi}\to{\to} \Liealg g \to \nul$ be
 a central extension. We choose an even linear map $\sigma: \Liealg g \to
\Liealg h$ satisfying $\pi \scirc \sigma = id(\Liealg g)$ (a section for $\pi$).
We then define $\Omega:\Liealg g \times \Liealg g \to E$ by
$$
i(\contrf{v,w}{\Omega}) = \contrf{\lieb{v,w}_{\Liealg g}}{\sigma} -
\lieb{\contrf{v}{\sigma}, \contrf{w}{\sigma}   }_{\Liealg h}
\mapob.
$$
And indeed the right hand side lies in the kernel of $\pi$, just because $\pi$
is a morphism of \glalg/s. Exactness of the sequence then shows that it is
indeed in the image of $i$; injectivity of $i$ then shows that
$\contrf{v,w}{\Omega} \in E$ is unique. Graded skew-symmetry of the Lie algebra
bracket shows that $\Omega$ is graded skew-symmetric, \ie, $\Omega \in
C^2(\Liealg g, E)$. The graded Jacobi identity translates to the fact that
$d\Omega = 0$~:
$$
\align
\contrf{u,v,w}{d\Omega}
&=
-\contrf{\lieb{u,v}, w}{\Omega}
+\altsign{\comme vw}{vw}
\contrf{\lieb{u,w},v}{\Omega}
+ \contrf{u, \lieb{v,w}}{\Omega}
\\&
=
-\contrf{\lieb{\lieb{u,v},w }}{\sigma} + \lieb{ \contrf{\lieb{u,v}}{\sigma},
\contrf{w}{\sigma} } 
\\&
\qquad
+\altsign{\comme vw}{vw}
\Bigl( \contrf{\lieb{\lieb{u,w},v }}{\sigma} - \lieb{
\contrf{\lieb{u,w}}{\sigma},
\contrf{v}{\sigma} }  \Bigr) 
\\&
\qquad
+ \contrf{\lieb{u, \lieb{v,w} }}{\sigma} - \lieb{ \contrf{u}{\sigma},
\contrf{\lieb{v,w}}{\sigma} }
\\&
=
\contrf{
-\lieb{\lieb{u,v},w }
+\altsign{\comme vw}{vw}
\lieb{\lieb{u,w},v }
+\lieb{u, \lieb{v,w} }
}{\sigma}
\\&
\qquad
+ 
\lieb{ \lieb{\contrf{u}{\sigma},\contrf{v}{\sigma}},
\contrf{w}{\sigma} } 
- \lieb{ \contrf{u}{\sigma}, \lieb{\contrf{v}{\sigma},\contrf{w}{\sigma}} }
\\&
\qquad
- \altsign{\comme vw}{vw}
\lieb{ \lieb{\contrf{u}{\sigma},\contrf{w}{\sigma}},
\contrf{v}{\sigma} } 
=0
\mapob,
\endalign
$$
where the last equality is a direct consequence of the graded Jacobi identity
in $\Liealg g$ and $\Liealg h$, and where the third equality is a direct
consequence of the definition of $\Omega$ and the fact that the extension is
central, meaning that all elements of the form $i(e)$, $e\in E$ disappear when
taking the bracket.

Once we know that $\Omega$ belongs to $Z^2(\Liealg g, E)$, we can form the
associated central extension $\Liealg g \times E$. 
Since $i(E)$ is the kernel of $\pi$ and since $\sigma$ is a section of $\pi$, the
map $\phi : \Liealg g \times E \to \Liealg h$,
$(v,e) \mapsto \contrf{v}{\sigma} - i(e)$ is an isomorphism of \gvs/s making
the diagram \recalf{equivofcentrext} commutative. To check that it is a
morphism of \glalg/s, we compute:
$$
\align
\contrf{ \lieb{ (v,e), (w,f) }}{\phi}
&=
\contrf{ ( \lieb{v,w}, \contrf{v,w}{\Omega} ) }{\phi}
=
\contrf{\lieb{v,w}}{\sigma} - i(\contrf{v,w}{\Omega})
\\&
=
\lieb{\contrf{v}{\sigma}, \contrf{w}{\sigma}   }
=
\lieb{\contrf{(v,e)}{\phi}, \contrf{(w,f)}{\phi}   }
\mapob,
\endalign
$$
where the last equality follows from the fact that $\Liealg h$ is a
\stress{central} extension of $\Liealg g$. We thus have shown that the
equivalence class of this extension is the same as the equivalence class
determined by $\Omega$, and thus we have proven surjectivity.
\QED\enddemo
} 

\heading \headnum. Central extensions and momentum maps \endheading

\proclaim{\thmm{symdefcentralext}{Proposition}}
Let $\Phi : G \times M \to M$ be the (left) action of an \glgrp/ $G$ on a
\ssmfd/ $(M,
\omega)$. For a fixed $m\in M$ we denote by $\Phi_m : G \to M$ the
map $g \mapsto \Phi(g, m)$ and we define $\omega\dom m =\Phi_m^*\omega $. 
If the $G$-action preserves $\omega$, then $\omega\dom m$ is a closed
left-invariant 2-form on $G$, \ie, $\double\omega\dom m \in Z^2(\Liealg g, C)$ If
$M$ is connected, the cohomology class of $\double\omega\dom m$ is independent of $m\in M$
and thus defines a unique  central extension of $\Liealg g$ by $C$
via the construction \recalt{centralextbycocycle}. 

\endproclaim

\dontprint{
\demo{Proof}
For $g\in G$ we have $L_g^* \omega\dom m = L_g^*(\Phi_m^* \omega) = (\Phi_m \scirc
L_g)^*\omega$. If we denote by $\Phi_g:M \to M$ the map $m\mapsto \Phi(g,m)$,
it follows immediately from the fact that $\Phi$ is a left action that $\Phi_m
\scirc L_g = \Phi_g \scirc \Phi_m$. Since the $G$-action preserves $\omega$ we
have $\Phi_g^*\omega = \omega$, and hence $L_g^* \omega\dom m = \omega\dom m$, \ie,
$\omega\dom m$ is a left-invariant 2-form on $G$. Since $d\omega\dom m = \Phi_m d\omega
= 0$, it is also closed. 

To prove that its cohomology class is independent of $m$, we consider the
function $\Omega : M \to Z^2(\Liealg g, C)$ defined by $\Omega(m) =
\double\omega\dom m$. For the projection $\pi \scirc \Omega : M \to H^2(\Liealg
g, C) = Z^2(\Liealg g, C)/B^2(\Liealg g, C)$ to be constant, it is necessary
and sufficient that $\forall m
\in M$
$\forall X_m \in T_mM$ we have
$X_m\Omega \in B^2(\Liealg g, C)$ (because $M$ is connected and $Z^2(\Liealg g,
C)$ finite dimensional). It follows directly from the definition of the
coboundary operator that $\lambda\in Z^2(\Liealg g, C)$ belongs to $B^2(\Liealg
g, C)$ if and only if there exists $\mu\in C^1(\Liealg g, C)$ such that
$\lambda = d\mu$,
\ie, $\forall v,w\in
\Liealg g : \contrf{v,w}{\lambda} =  \contrf{\lieb{v,w}}{\mu}$ ($v$ and $w$
are constants with respect to a derivation in the direction of $m$). 
We thus want to compute
$\contrf{v,w}{X_m\Omega}$. For homogeneous $X$, $v$, and $w$, this is
the same as $(-1)^{\e(X)(\e(v) + \e(w))} X_m\bigl( \contrf{v,w}{\Omega}
\bigr)$. We now compute:
$$
\contrf{v,w}{\Omega(m)} = \contrf{v,w}{\double\omega\dom m} = \contrf{-v^M,
-w^M}{\double\omega_m}
\mapob,
$$
and thus $d\contrf{v,w}{\Omega} = \contrf{\lieb{v,w}^M}{\double\omega}$ by
\recalt{commlocisglob}. We thus find (for arbitrary $v$, $w$, $X_m$) 
$$
\contrf{v,w}{X_m\Omega} = -\contrf{\lieb{v,w}^M, X_m}{\double\omega_m}
\mapob.
$$
Hence if we define $\mu \in C^1(\Liealg g, C)$ by $\contrf{u}{\mu} = -
\contrf{u^M, X_m}{\double\omega_m}$, then $X_m\Omega = d\mu$, \ie,
$X_m\Omega \in B^2(\Liealg g, C)$ as wanted. This proves that
for connected $M$ the cohomology class of $\Omega(m)$ is independent of $m\in M$.
\QED\enddemo
} 

\definition{\thm{Definitions}}
Let $G$ be an \glgrp/ with \glalg/ $\Liealg g = T_eG$ and suppose that $G$ acts
on a \ssmfd/ $(M, \omega)$. If the action preserves the symplectic form
$\omega$,  it also preserves the homogeneous parts $\omega_\alpha$
separately because diffeomorphisms are even.  If the
$G$-action preserves $\omega_\alpha$, then the Lie derivative of $\omega_\alpha$
in the direction of a fundamental vector is zero, \ie, $v\in \Liealg g
\Rightarrow \Lied(v^M)\omega_\alpha = 0$. It follows that the fundamental vector
fields associated to the $G$-action are locally hamiltonian. 
According to
\recalt{commlocisglob} we have $\contrf{\lieb{x,y}^M}{\double\omega} =
d\contrf{x^M, y^M}{\double\omega}$, from which it follows that for all $z\in
\lieb{\Liealg g,
\Liealg g}$ (the commutator subalgebra) the fundamental vector field $z^M$ is
globally hamiltonian.
The $G$-action is called \stress{(weakly) hamiltonian} if \stress{all}
fundamental vector fields are globally hamiltonian. 

As for $k$-forms, we can transform any map $J:M \to \Liealg g^*$ into an
\stress{even} map $\double J:M \to \Liealg g^* \otimes C$ by $\double J = 
J_0 \otimes c_0 + J_1 \otimes
c_1$,  where $J_\alpha : M \to \Liealg g^*_\alpha$ denotes the homogeneous
component of parity $\alpha$ of $J$. A map $J:M\to \Liealg g^*$ is called a
\stress{momentum map} for a weakly hamiltonian action if 
$$
\forall v \in \Liealg g : \contrf{v^M}{\double\omega} = d\contrs{v}{\double J}
\mapob,
$$
which is equivalent to the condition $
\forall v\in \Liealg g, \forall \alpha = 0,1 : 
\contrf{v^M}{\omega_\alpha} = d\contrs{v}{J_\alpha}
$.
In terms of hamiltonian vector fields we can interpret a momentum map as a map
$\Liealg g \to \Poisson$, $v\mapsto \contrs{v}{\double J}$ satisfying the condition
$$
\forall v\in \Liealg g : v^M = X_{\contrs{v}{\double J}}
\mapob.
$$

The $G$-action is called \stress{strongly hamiltonian} if there exists  an
equivariant momentum map, which means in our context that the map $\Liealg g
\to \Poisson$ is a morphism of \glalg/s:
$$
\forall v,w \in \Liealg g :
\{\ \contrs{v}{\double J}
\ ,\ 
\contrs{w}{\double J}
\ \}
=
\contrs{\lieb{v,w}}{\double J}
\mapob.
$$

Using the fact that a momentum map always exists on the commutator subalgebra,
one can give an alternative proof of \recalt{symdefcentralext}.

\enddefinition

\dontprint{
\demo{Alternative proof of \recalt{symdefcentralext}}
Let $\Liealg s = \lieb{\Liealg g, \Liealg g} \subset \Liealg g$ denote the
commutator subalgebra and let $\Liealg t \subset \Liealg g$ be a supplement for
$\Liealg s$, \ie, $\Liealg g = \Liealg s \oplus \Liealg t$. We know that for
all $z\in \Liealg s$ the associated fundamental vector field $z^M$ is globally
hamiltonian, and thus in particular there exists a smooth map $J: M \to \Liealg
s^*$ such that for all $z\in \Liealg s$ we have $z^M = X_{\contrs{z}{\double J}}$. 
Moreover, for $x,y \in \Liealg g$ we have
$$
d\contrf{x^M, y^M}{\double\omega} = \contrf{\lieb{x,y}^M}{\double\omega} = d\contrs{
\lieb{x,y}}{\double J}
\mapob.
$$
Since $M$ is connected, this implies that the function $\contrf{x^M,
y^M}{\double\omega} - \contrs{ \lieb{x,y}}{\double J}$ is constant. We now define $\mu
\in C^1(\Liealg g, C)$ by
$$
\contrs{u}{\mu} = \cases
0 & u\in \Liealg t
\\
\contrs{u}{\double J}(m') - \contrs{u}{\double J}(m)  & u\in \Liealg s
\mapob.
\endcases
$$
We now compute
$$
\align
\contrf{x,y}{\double\omega\dom m} - \contrf{x,y}{\double\omega\dom{m'}}
&=
\contrf{x^M, y^M}{\double\omega_m} - \contrf{x^M, y^M}{\double\omega_{m'}}
\\
&=
\contrs{ \lieb{x,y}}{\double J}(m) - \contrs{ \lieb{x,y}}{\double J}(m') 
=
-\contrs{ \lieb{x,y}}{\mu}
\mapob.
\endalign
$$
Hence $\double\omega\dom m - \double\omega\dom{m'} = d\mu$, \ie,
$\double\omega\dom{m}$ and
$\double\omega\dom{m'}$ determine the same element in $H^2(\Liealg g, C)$.
\QED\enddemo
} 

\proclaim{\thmm{cocyclebymomentum}{Lemma}}
Let $G$ be a symmetry group of a connected \ssmfd/ $(M, \omega)$ admitting a
momentum map $J$ (\ie, the action is hamiltonian). Then
\roster
\item"(i)"
for all
$v,w \in \Liealg g$ the function $\contrf{v,w}{\Omega_J} : M \to C$, 
$$\contrf{v,w}{\Omega_J} : m \mapsto
\PB{
\contrs{v}{\double J}, \contrs{w}{\double J}
}(m) - \contrs{ \lieb{v,w} }{\double J(m)} 
$$ 
is a constant function;

\item"(ii)"
the cohomology class of $\double\omega\dom m$ defined in
\recalt{symdefcentralext} is also determined by the 2-cocycle $(v,w) \mapsto
\contrf{v,w}{\Omega_J}$ given by (i).

\endroster

\endproclaim

\dontprint{
\demo{Proof}
By definition of the Poisson bracket we have $\{ \contrs{v}{\double J},
\contrs{w}{\double J}
\} = \contrf{v^M, w^M}{\double\omega}$. By \recalt{commlocisglob} the
exterior derivative of this function is
$\contrf{\lieb{v,w}^M}{\double\omega}$, which is equal to
$d\contrs{\lieb{v,w}}{\double J}$ by definition of the momentum map. This proves
(i).

In the proof of \recalt{symdefcentralext} we have seen the equalities
$\contrf{v,w}{\double\omega\dom m} = 
\contrf{v^M,w^M}{\double\omega_m} = \{ \contrs{v}{\double J},
\contrs{w}{\double J} \}(m) $. For a fixed $m\in M$ we can see $\double J(m)$ as an element
of $C^1(\Liealg g, C)$ and then 
$$
\contrf{v,w}{\double\omega\dom m} = 
\{ \contrs{v}{\double J},
\contrs{w}{\double J} \}(m) 
=
\contrf{v,w}{\Omega_J} + \contrs{\lieb{v,w}}{\double J(m)}
\mapob,
$$
and thus $\double\omega\dom m = \Omega_J + d (\double J(m))$, where here the $d$
denotes the coboundary operator $C^1 \to B^2 \subset Z^2$. This proves that the
cohomology classes of $\double\omega\dom m$ and $\Omega_J$ are the same.
\QED\enddemo
} 

\remark{Remark}
Note that we never explicitly proved that $\Omega_J$ is an element of
$Z^2(\Liealg g, C)$. However, the fact that we have the equality $\Omega_J =
\double\omega\dom m + d \double J(m)$ implies automatically that it indeed is a
2-cocycle.

\endremark

\heading \headnum. Coadjoint orbits \endheading

If $G$ is an \glgrp/ of dimension $p|q$, its Lie algebra
$\Liealg g$, as well as its (left) dual $\Liealg g^*$ is \agvs/ of dimension
$p|q$, meaning that there is a basis with $p$ even vectors and $q$ odd vectors.
But since the coordinates belong to the full ring $\CA$, the dimension of
$\Liealg g^*$ seen as an \gmfd/ is $n|n$, where $n=p+q$. What we are going
to show is that the coadjoint orbit $\Orbit_\mu$ through a point $\mu \in
\body \Liealg g^*$ (\ie, $\mu$ has real coordinates with respect to the basis)
has a natural symplectic form. As we will see in an explicit example, this
symplectic form need not be homogeneous, nor need it be non-degenerate.

Let $e_1, \dots, e_n$ be a homogeneous basis for $\Liealg g$, and let $e^1,
\dots, e^n$ be the (left) dual basis for $\Liealg g^*$. In $\Liealg g^*$ we
introduce  $2n$ homogeneous coordinates $\mu_1, \dots, \mu_n, \mub_1, \dots,
\mub_n$ with parities
$\e(\mu_i) = \e(e^i)$, $\e(\mub_i) = 1-\e(e^i)$ of a point $\mu \in
\Liealg g^*$ according to the formula
$$
\mu = \sum_{i = 1}^n (\mu_i + \mub_i) \cdot e^i
\mapob.
$$
In order to distinguish the subscript indicating the homogeneous part of a
vector from the subscript indicating the coordinate, we will put parentheses
around a vector before taking the homogeneous parts. More precisely,
$(\mu)_\alpha$ denotes the homogeneous part of parity $\alpha$ of the vector
$\mu$. In terms of the coordinates introduced above we have
$$
(\mu)_0 = \sum_{i = 1}^n \mu_i \cdot e^i
\qquad\text{and}\qquad
(\mu)_1 = \sum_{i = 1}^n \mub_i \cdot e^i
\mapob.
$$
From these equations we learn that we can define the coordinates of $\mu$ also
as
$$
\mu_i = (-1)^{\e(e_i)} \contrs{e_i}{(\mu)_0}
\qquad\text{and}\qquad
\mub_i = \contrs{e_i}{(\mu)_1}
\mapob.
$$

The coadjoint action of $G$ on $\mu \in \Liealg g^*$ is defined via the adjoint
representation according to the formula $\forall v\in
\Liealg g : \contrs{v}{\Coad(g)\mu} =
\contrs{\Ad(g\mo)v}{\mu}$. Just as the algebraic adjoint representation $\ad$ of
$\Liealg g$ is the infinitesimal version of $\Ad$ in the sense $\ad = T_e\Ad$,
so is the algebraic coadjoint representation $\coad$ the infinitesimal version
of $\Coad\,$: $\coad = T_e\Coad : \Liealg g \to \End_R(\Liealg g^*)$. More
precisely, let $g^1, \dots, g^n$ be local coordinates on $G$ such that the
tangent vectors $\partial_{g^i}\restricted_e$ are the basis vectors $e_i$ of
$\Liealg g = T_eG$. Then $\sum_{i=1}^n w^i \partial_{g^i}\restricted_e
\Coad(g) = \coad(\sum_{i=1}^n w^i e_i)$.

\proclaim{\thmm{Coaddercoad}{Lemma}}
For homogeneous $v,w\in \Liealg g$ we have
$$
\align
\sum_{i=1}^n w^i \fracp{}{g^i}\restricted_e \contrs{v}{\Coad(g)\mu} 
&=
\comme vw
\contrs{v}{\coad (w) \mu} 
\\
&= 
\comme vw
\contrs{\lieb{v,w }}{\mu} = - \contrs{\lieb{w,v}}{\mu}
\mapob.
\endalign
$$ 
In particular, $\forall v,w \in \Liealg g :
\contrs{v}{\coad(w)\mu} = \contrs{\lieb{v,w}}{\mu}$.

\endproclaim

\dontprint{
\demo{Proof}
Using the defining
equation $\contrs{v}{\Coad(g)\mu} = \contrs{\Ad(g\mo)v}{\mu}$ we compute
$$
\align
\sum_{i=1}^n w^i \fracp{}{g^i}\restricted_e \contrs{v}{\Coad(g)\mu} 
&=
\sum_{i=1}^n w^i \fracp{}{g^i}\restricted_e \contrs{\Ad(g\mo)v}{\mu}
\\
\comme vw
\contrs{v}{\sum_{i=1}^n w^i \fracp{}{g^i}\restricted_e
\Coad(g)\mu}
&=
-\contrs{\ad(w)v}{\mu} = -\contrs{\lieb{w,v}}{\mu}
\\
\comme vw
\contrs{v}{\coad(w)\mu}
&=
\comme vw \contrs{\lieb{v,w}}{\mu}
\mapob.
\endalign
$$
The particular case follows directly by replacing $w$ by $\comme vw w$.
\QED\enddemo
} 

In order to determine the action of $\Coad(g)$ in terms of our coordinates on
$\Liealg g^*$, we first note that the matrix elements of  $\Coad(g)$ are
defined  by $\Coad(g)_i{}^j = \contrs{e_i}{\Coad(g)e^j}$, which is equivalent
to saying that $\Coad(g)e^j = \sum_j e^i \Coad(g)_i{}^j$. Similarly, the matrix
elements of $\coad$ are defined by $\coad(w)_i{}^j =
\contrs{e_i}{\coad(w)e^j}$. Since $\Coad(g)$ is an even map it follows that
the parity of these matrix elements is given as
$\e(\Coad(g)_i{}^j) = \e(e_i) + \e(e_j)$. Either from these parity
considerations, or from the fact that $\Coad(g)$ is even and thus preserves the
parity decomposition in $\Liealg g^*$, \ie, $(\Coad(g)\mu)_\alpha = \Coad(g)
(\mu)_\alpha$, one can deduce that the action of $\Coad(g)$ is given in terms of
coordinates as
$$
\align
\mu_i &\mapsto \sum_{j=1}^n \comm{\e(e_i)}{(\e(e_i) + \e(e_j))} \mu_j
\Coad(g)_i{}^j =  (-1)^{\e(e_i)} \contrs{e_i}{\Coad(g)(\mu)_0}
\mapob,
\\
\mub_i &\mapsto \sum_{j=1}^n \comm{\e(e_i)}{(\e(e_i) + \e(e_j))} \mub_j
\Coad(g)_i{}^j = \contrs{e_i}{\Coad(g)(\mu)_1}
\mapob.
\endalign
$$

In order to compute the fundamental vector field $v^{\Liealg g^*}$ on $\Liealg
g^*$ associated to a vector $v\in\Liealg g$, we recall that it is defined
 as $v^{\Liealg g^*}\restricted_\mu = -\sum_{k=1}^n v^k
\partial_{g^k}\restricted_e \Coad(g)\mu$. Applying this to the coordinate
expression of $\Coad(g)\mu$ and using \recalt{Coaddercoad} to compute the
derivative of matrix elements, we obtain the following result.

\proclaim{\thmm{fundvectfield}{Lemma}}
The tangent vector $v^{\Liealg g^*}\restricted_\mu$  is given by
$$
- \sum_{i=1}^n \Bigl\{
\invol^{\e(e_i)}\Bigl( \contrs{e_i}{\coad(v)(\mu)_0}   \Bigr)
\cdot
\fracp{}{\mu_i}\restricted_\mu
+
\invol^{\e(e_i)}\Bigl( \contrs{e_i}{\coad(v)(\mu)_1}   \Bigr)
\cdot
\fracp{}{\mub_i}\restricted_\mu
\Bigr\}
\mapob.
$$
In particular, $v^{\Liealg g^*}\restricted_\mu$ is zero if and
only if $\forall\alpha : \coad(v)(\mu)_\alpha = 0$. 

\endproclaim

\dontprint{
\demo{Proof}
\vskip-3\baselineskip
$$
\align
v^{\Liealg g^*}\restricted_\mu 
&= 
- \sum_{i,k=1}^n (-1)^{\e(e_i)} 
v^k \fracp{}{g^k}\restricted_e\bigl( \contrs{e_i}{(\Coad(g)(\mu)_0)} \bigr)
\cdot\fracp{}{\mu_i}\restricted_\mu
\\ 
&\qquad\qquad
- \sum_{i,k=1}^n
v^k \fracp{}{g^k}\restricted_e\bigl( \contrs{e_i}{(\Coad(g)(\mu)_1)} \bigr)
\cdot\fracp{}{\mub_i}\restricted_\mu
\\
&=
- \sum_{i,k=1}^n (-1)^{\e(e_i)(\e(e_i)+\e(e_k))} 
v^k \contrs{e_i}{(\coad(e_k)(\mu)_0)} \cdot\fracp{}{\mu_i}\restricted_\mu
\\
&\qquad\qquad
- \sum_{i,k=1}^n \comme{e_i}{e_k} v^k 
\contrs{e_i}{(\coad(e_k)(\mu)_1)} \cdot\fracp{}{\mub_i}\restricted_\mu
\\
&=
- \sum_{i=1}^n 
\invol^{\e(e_i)}\Bigl( \contrs{e_i}{(\coad(v)(\mu)_0)}   \Bigr)
\cdot
\fracp{}{\mu_i}\restricted_\mu
\\
&\qquad\qquad
- \sum_{i=1}^n 
\invol^{\e(e_i)}\Bigl( \contrs{e_i}{(\coad(v)(\mu)_1)}   \Bigr)
\cdot
\fracp{}{\mub_i}\restricted_\mu
\mapob.
\endalign
$$
Since $\sum_{i=1}^n e^i x_i = \sum_{i=1}^n \invol^{\e(e_i)}\bigl(x_i\bigr)
e^i$, this means that the coefficients of $\partial_{\mu_i}$ and
$\partial_{\mub_i}$ are the left coordinates of $\coad(v)(\mu)_0$ and
$\coad(v)(\mu)_1$ respectively.  
\QED\enddemo
} 

We now have sufficient material to define the symplectic form $\omega$ on a
coadjoint orbit $\Orbit_{\mu_o} = \{ \Coad(g){\mu_o} \mid g\in G\}$. Since we
choose ${\mu_o} \in \body \Liealg g^*$, this is indeed \agmfd/, immersed in
$\Liealg g^*$. The form $\omega$ is also called the Kirillov-Kostant-Souriau
form. For any $\mu\in \Orbit_{\mu_o}$, the tangent space $T_\mu\Orbit_{\mu_o}$
is given by the set of all fundamental vector fields at $\mu\,$:
$$
T_\mu\Orbit_{\mu_o} = \{ v^{\Liealg g^*}\restricted_\mu \mid v \in \Liealg g \}
\mapob.
$$
We then define $\omega$ by its action on tangent vectors by
$$
\contrf{v^{\Liealg g^*}\restricted_\mu, 
w^{\Liealg g^*}\restricted_\mu}{\omega\restricted_\mu} =
\contrs{\lieb{v,w}}{\mu} \equiv \contrs{v}{\coad(w)\mu}
\mapob.
$$

\proclaim{\thm{Lemma}}
$\omega$ is a well defined closed and homogeneously non-degenerate 2-form on
$\Orbit_{\mu_o}$, \ie, $\Orbit_{\mu_o}$ is a \ssmfd/. Moreover, the coadjoint
action is strongly hamiltonian with momentum map $J(\mu) = \mu$.

\endproclaim

\dontprint{
\demo{Proof}
From the equality $\contrf{v^{\Liealg g^*}\restricted_\mu, 
w^{\Liealg g^*}\restricted_\mu}{\omega\restricted_\mu} =
\contrs{\lieb{v,w}}{\mu}$ we see immediately that $\omega\restricted_\mu$ is
graded skew-symmetric, and from  $\contrf{v^{\Liealg g^*}\restricted_\mu, 
w^{\Liealg g^*}\restricted_\mu}{\omega\restricted_\mu} =
\contrs{v}{\coad(w)\mu}$ and \recalt{fundvectfield} we see that if $w^{\Liealg
g^*}\restricted_\mu=0$, then $\coad(w)\mu=0$ and thus 
$\contrf{v^{\Liealg g^*}\restricted_\mu,  w^{\Liealg
g^*}\restricted_\mu}{\omega\restricted_\mu} = 0$. In other words, 
$\contrf{v^{\Liealg g^*}\restricted_\mu,  w^{\Liealg
g^*}\restricted_\mu}{\omega\restricted_\mu}$  is independent of the choice of
$w$ as long as $w^{\Liealg g^*}\restricted_\mu$ does not change. Combining this
with the graded skew-symmetry, we see that
$\omega\restricted_\mu$ is a well defined graded skew-symmetric form on
$T_\mu\Orbit_{\mu_o}$.

To show that it is homogeneously non-degenerate, we first note that the
definitions of $\omega$ and $J$ directly gives the homogeneous parts
$\omega_\alpha$ as
$$
\align
\contrf{v^{\Liealg g^*}\restricted_\mu, 
w^{\Liealg g^*}\restricted_\mu}{\omega_\alpha\restricted_\mu} =
\contrs{\lieb{v,w}}{(\mu)_\alpha} &= \contrs{v}{\coad(w)(\mu)_\alpha}
\formula{firstorbitomega}
\\
&= \contrs{ \lieb{v,w} }{J_\alpha(\mu)}
\mapob.
\formula{secondorbitomega}
\endalign
$$
Now suppose that $w\in \Liealg g$ is such that $\forall \alpha : 
\contrf{w^{\Liealg
g^*}\restricted_\mu}{\omega_\alpha\restricted_\mu} = 0$, \ie, $\forall \alpha, 
\forall v\in \Liealg g$~: $ \contrf{v^{\Liealg g^*}\restricted_\mu,  w^{\Liealg
g^*}\restricted_\mu}{\omega_\alpha\restricted_\mu} = 0$. Formula
\recalf{firstorbitomega} directly implies that $\forall \alpha :
\coad(w)(\mu)_\alpha = 0$, and thus
$w^{\Liealg g^*}\restricted_\mu = 0$ according to \recalt{fundvectfield}. This
proves that $\omega$ is homogeneously non-degenerate.

To show that $\omega$ is closed and $J$ strongly hamiltonian, we make a
preliminary computation. We first note that
for $v= \sum_{i=1}^n v^i e_i$ and $\mu = \sum_{i=1}^n (\mu_i + \mub_i)e^i$ we
obtain $\contrs{v}{\mu} = \sum_{i=1}^n v^i ( (-1)^{\e(e_i)} \mu_i + \mub_i)$.
Using the explicit expression for $u^{\Liealg g^*}$
given in \recalt{fundvectfield} we compute:
$$
\gather
u^{\Liealg g^*}\restricted_\mu \contrs{ v}{J}   
\\
{\align
&=
- \sum_{i,j,k=1}^n \Bigl(
(-1)^{\e(e_i)(\e(e_i)+\e(e_k))} 
u^k \contrs{e_i}{\coad(e_k)(\mu)_0} \cdot\fracp{}{\mu_i}\restricted_\mu
\\
&\qquad\qquad
+ \comme{e_i}{e_k} u^k 
\contrs{e_i}{\coad(e_k)(\mu)_1} \cdot\fracp{}{\mub_i}\restricted_\mu
\Bigr)
v^j ( (-1)^{\e(e_j)} \mu_j + \mub_j)
\\
&=
- \sum_{i,k=1}^n \Bigl(
(-1)^{\e(e_i)(\e(e_i)+\e(e_k))} 
u^k \contrs{e_i}{\coad(e_k)(\mu)_0} 
v^i (-1)^{\e(e_i)(\e(v)+\e(w))}
\\
&\qquad\qquad
+\comme{e_i}{e_k} u^k 
\contrs{e_i}{\coad(e_k)(\mu)_1}
v^i (-1)^{(\e(e_i)+1)(\e(v)+\e(w))}
\Bigr)
\\
&=
- \sum_{k=1}^n 
(-1)^{\e(e_k)(\e(v)+\e(w))} 
u^k \contrs{v}{\coad(e_k)((\mu)_0 + (\mu)_1) } 
\\
&=
- (-1)^{\e(u)(\e(v)+\e(w))}
\contrs{v}{\coad(u)\mu}
= \contrs{\lieb{u, v}}{\mu}
\mapob.
\endalign
}
\endgather
$$
tracing what happens with $u^{\Liealg g^*}\restricted_\mu \contrs{
v}{J_\alpha}$ we find
$$
u^{\Liealg g^*}\restricted_\mu \contrs{ v}{J_\alpha}
=
\contrs{\lieb{u, v}}{J_\alpha(\mu)}
\mapob.
\formula{interorbitomega}
$$

Combining \recalf{interorbitomega} with \recalf{secondorbitomega} we
immediately have $d\contrs{v}{J_\alpha} = \contrf{v^{\Liealg
g^*}}{\omega_\alpha}$,
\ie, the action is (weakly) hamiltonian. Once we know that $v^{\Liealg g^*}$ is
the hamiltonian vector field associated to $\contrs{v}{\double J}$,
we can combine \recalf{interorbitomega} with the
definition of the Poisson bracket to obtain that the action is strongly
hamiltonian.

For the last item on our list, $\omega$ closed, we evaluate
$d\omega$ on three homogeneous vector fields according to
\recalf{extdertwoform}.
Since the map $v\mapsto v^{\Liealg g^*}$ is a morphism of graded Lie algebras,
the terms like $\contrf{\lieb{u^{\Liealg g^*}, v^{\Liealg g^*}}, w^{\Liealg
g^*}}{\omega\restricted_\mu}$ become $\contrs{\lieb{ \lieb{u, v}, w}}{\mu}$. 
Combining \recalf{secondorbitomega} with \recalf{interorbitomega} shows that
the terms like $u^{\Liealg g^*}\restricted_\mu\big( \contrf{ v^{\Liealg g^*}, 
w^{\Liealg g^*}}{\omega}   \bigr)$ are also of the form $\contrs{\lieb{
\lieb{u, v}, w}}{\mu}$.
We thus find: 
$$
\align
&\hskip4cm
\contrf{u^{\Liealg g^*}, v^{\Liealg g^*}, 
w^{\Liealg g^*}}{d\omega\restricted_\mu}
=
\\
\vspace{2\jot}
&=
u^{\Liealg g^*}\restricted_\mu\big( \contrf{ v^{\Liealg g^*}, 
w^{\Liealg g^*}}{\omega}   \bigr)
-\comme uv v^{\Liealg g^*}\restricted_\mu\big( \contrf{ u^{\Liealg g^*}, 
w^{\Liealg g^*}}{\omega}   \bigr)
\\
&\qquad\qquad
+ (-1)^{\e(w)(\e(u)+\e(v))} w^{\Liealg g^*}\restricted_\mu\big( \contrf{
u^{\Liealg g^*},  v^{\Liealg g^*}}{\omega}   \bigr)
- \contrf{\lieb{u^{\Liealg g^*}, v^{\Liealg g^*}}, w^{\Liealg
g^*}}{\omega\restricted_\mu}
\\
&\qquad\qquad
+ \comme vw \contrf{\lieb{u^{\Liealg g^*}, w^{\Liealg g^*}}, v^{\Liealg
g^*}}{\omega\restricted_\mu}
+ \contrf{u^{\Liealg g^*}, \lieb{v^{\Liealg g^*}, w^{\Liealg
g^*}}}{\omega\restricted_\mu}
\\
\allowdisplaybreak
&=
\contrf{\lieb{u, \lieb{v,w}}}{\mu} 
-\comme uv \contrf{\lieb{v, \lieb{u,w}}}{\mu}
+ (-1)^{\e(w)(\e(u)+\e(v))} \contrf{\lieb{w, \lieb{u,v}}}{\mu}
\\
&\qquad\qquad
- \contrf{\lieb{\lieb{u,v},w}}{\mu}
+ \comme vw \contrf{\lieb{\lieb{u,w},v}}{\mu}
+ \contrf{\lieb{u, \lieb{v,w}}}{\mu}
\\
\allowdisplaybreak
&=
2 \comme uw 
\Bigl( \comme uw \contrf{\lieb{u, \lieb{v,w}}}{\mu} 
+ \comme uv \contrf{\lieb{v, \lieb{w,u}}}{\mu}
\\
&\hskip6em
+ \comme wv \contrf{\lieb{w, \lieb{u,v}}}{\mu}
\Bigr)
= 0 \qquad\text{by the Jacobi identity.}
\endalign
$$
This proves that $d\omega$ evaluated on three homogeneous vectors is always
zero. But then by trilinearity $d\omega$ is zero.
\QED\enddemo
} 

\heading \memorize\headnum={Heisenbergsec}. Super Heisenberg groups \endheading

\subheading{ The general case }

Let $E$ be \agvs/ of dimension $p|q$ with homogeneous basis $e_1, \dots,
e_n$, $n=p+q$ %
 and let $\Omega: E \times E \to C$ be an even graded
skew-symmetric bilinear form. The (left) coordinates of the element $\contrf{v, \hat v}\Omega \in C$ with
respect to the basis $c_0,c_1$ determine an even and an odd graded
skew-symmetric bilinear form $\Omega^0$ and $\Omega^1$ on
$E$ with values in $\CA$ by the formula 
$\contrf{v,\hat v}\Omega = \contrf{v,\hat v}{\Omega^0}\cdot c_0 + \contrf{v,\hat
v}{\Omega^1} \cdot c_1$.
With these ingredients we define the \glgrp/ $G$
as follows. As set
$G$ is $(E \times C)_0$, and the group structure is given by
$$
(a,b) \cdot (\hat a, \hat b) = (a+\hat a, b+ \hat b + \tfrac12 \contrf{a,\hat
a}\Omega)
\mapob.
$$
The neutral element is $(0,0)$, and the inverse of $(a,b)$ is $(a,b)\mo = (-a,
-b)$.

On $G$ we introduce $n+2$ homogeneous coordinates $a^1, \cdots, a^n, b^0, b^1$
according to the formula $(a,b) = (\sum_i a^i e_i,
\sum_\alpha b^\alpha c_\alpha)$. Their parities are given by $\e(a^i) = \e(e_i)$
and
$\e(b^\alpha)  = \e(c_\alpha)$. In terms of these coordinates a basis of the
left invariant vector fields is given as
$$
\fracp{}{a^i}\restricted_{(a,b)} + \tfrac12 \contrf{a,e_i}{\Omega^\alpha}
\fracp{}{b^\alpha}\restricted_{(a,b)}
\qquad\text{and}\qquad
\fracp{}{b^\alpha}\restricted_{(a,b)}
\mapob.
$$
Identifying the tangent space $T_{(0,0)}G$ with $E \times C$, we can identify
the above basis for the \glalg/ $\Liealg g$ with the basis for $E \times C$, and
in particular we can identify $e_i$ with the left-invariant vector field
$\partial_{a^i}\restricted_{(a,b)} + \tfrac12 \contrf{a,e_i}{\Omega^\alpha}
\partial_{b^\alpha}\restricted_{(a,b)}$ and $c_\alpha$ with
$\partial_{b^\alpha}\restricted_{(a,b)}$.  The only non-zero commutators among
these left-invariant vector fields are 
$$
\multline
\Bigl[\ \fracp{}{a^i}\restricted_{(a,b)} + \tfrac12
\contrf{a,e_i}{\Omega^\alpha}
\fracp{}{b^\alpha}\restricted_{(a,b)}
\ ,\ 
\fracp{}{a^j}\restricted_{(a,b)} + \tfrac12 \contrf{a,e_j}{\Omega^\beta}
\fracp{}{b^\beta}\restricted_{(a,b)}
\ \Bigr] = 
\dontprint{
\\
\Bigl( \tfrac12 \contrf{e_i,e_j}{\Omega^\gamma} - \tfrac12 \comme{e_i}{e_j}
\contrf{e_j,e_i}{\Omega^\gamma}
\Bigr)
\fracp{}{b^\gamma}\restricted_{(a,b)}
}  
\\
= \contrf{e_i,e_j}{\Omega^\gamma} \fracp{}{b^\gamma}\restricted_{(a,b)}
\mapob,
\endmultline
$$
in other words, $[e_i,e_j] = \contrf{e_i,e_j}{\Omega^\alpha} c_\alpha$. For
general $(v,z), (\vh, \zh) \in E \times C = \Liealg g$ this gives the
commutator $\lieb{(v,z), (\vh, \zh)} = (0, \contrf{v,\vh}{\Omega})$.

For $g=(a,b) \in G$ and $h=(\hat a, \hat b)$ we find $ghg\mo = (\hat a, \hat
b + \contrf{a,\hat a}\Omega)$. This gives us for the Adjoint representation the
formul\ae{}
$$
\Ad(g) e_i = e_i + \contrf{a,e_i}{\Omega^\alpha} c_\alpha
\qquad\text{and}\qquad
\Ad(g)c_\alpha = c_\alpha
\mapob.
$$
We now consider the left dual algebra $\Liealg g^* = E^* \times C^*$, \ie, the
space of all left linear maps from $\Liealg g$ to $\CA$ and we denote the dual
basis by
$e^1,
\cdots, e^n, c^0, c^1$.
By definition of the coadjoint representation, we thus have
the equalities
$\contrs{e_i}{\Coad(g)e^j} = \delta_i^j$, $\contrs{c_\alpha}{\Coad(g)e^j} = 0$, 
$\contrs{e_i}{\Coad(g)c^\beta} = -\contrf{v,e_i}{\Omega^\beta}$,
$\contrs{c_\alpha}{\Coad(g)c^\beta} = \delta_\alpha^\beta$, from which we deduce
that the coadjoint representation
is given by the formul\ae{}
$$
\Coad(g)e^i = e^i
\qquad\text{and}\qquad
\Coad(g) c^\alpha = c^\alpha + e^i \cdot \contrf{e_i,a}{\Omega^\alpha}
= c^\alpha + \contrf{a}{\Omega^\alpha}
\mapob,
$$
where $\contrf{a}{\Omega^\alpha} \in E^*$ denotes the left linear map $v
\mapsto \contrf{v,a}{\Omega^\alpha}$.

When introducing coordinates on $\Liealg g^*$, we should remember that we
look at the full dual and thus that the coefficient of a basis vector takes its
values in $\CA$. We thus introduce $2n+4$ homogeneous coordinates  $x_1,
\dots, x_n, \xb_1, \dots, \xb_n, y_0, y_1, \yb_0, \yb_1$ with parities
$\e(x_i) = \e(e^i)$, $\e(\xb_i) =
 1-\e(e^i)$, $\e(y_\alpha) = \e(c_\alpha)$, $\e(\yb_\alpha) = 1-\e(c_\alpha)$ of
a point
$\mu = (x, y) \in \Liealg g^*$ according to the formula
$$
(x, y) = (y_0 +\yb_0)\cdot c^0 + (y_1 +\yb_0)\cdot c^1 + \sum_{i =
1}^n (x_i + \xb_i) \cdot e^i 
\mapob.
$$
It follows that the coadjoint action of $g = (a,b) \in G$
on an element $(x, y) \in \Liealg g^*$ is given by
$$
\Coad(g)(x, y) = (x + (y_0 + \yb_0) \cdot \contrf{a}{\Omega^0} +
(y_1 + \yb_1)  \cdot
\contrf{a}{\Omega^1}\ ,\ y)
\mapob.
$$

In order to have genuine submanifolds, we now specialize to the case of an orbit
through a point with real coordinates (and thus in particular $y_1 = \yb_0 =
0$). The $y$-coordinates do not change under the action of $\Coad(g)\,$; the
$x$-coordinates change according to 
$$
x_i \mapsto x_i - (-1)^{\e(e_i)}\, y_0 \cdot \contrf{a,e_i}{\Omega^0}
\qquad\text{and}\qquad
\xb_i \mapsto \xb_i - \yb_1\cdot \contrf{a,e_i}{\Omega^1}
\mapob,
$$
where the sign $(-1)^{\e(e_i)}$ comes from interchanging $e^i$ and
$\contrf{a,e_i}{\Omega^\alpha}$. 
It follows that there are three different types of orbit depending on whether
$y_0$  or $\yb_1$ is zero, their dimension depending upon the
dimension of the image of the maps $\Omega^\alpha : E \to E^*\,$; the fourth case
$y_0 = \yb_1 = 0$  yields the trivial orbit $\nul$ of dimension~$0|0$. 

Since the action is linear, it is straightforward to compute the fundamental
vector field
$(v,z)^{\Liealg g^*}$ 
associated to the element $(v,z) \in \Liealg g\,$:
$$
(v,z)^{\Liealg g^*}\restricted_{(x,y)} = 
 \sum_i \Bigl(
(-1)^{\e(e_i)}\, y_0 \cdot \contrf{v,e_i}{\Omega^0} \cdot
\fracp{}{x_i}\restricted_{(x,y)} 
+ \yb_1\cdot \contrf{v,e_i}{\Omega^1} \cdot 
\fracp{}{\xb_i}\restricted_{(x,y)} 
\Bigr)
\mapob.
$$
For the symplectic form $\omega$ on an orbit we obtain
the formula
$$
\contrf{(v,z)^{\Liealg g^*}, (\hat v, \hat z)^{\Liealg g^*}}{\omega_{(x,y)}}
= 
\contrf{\ \bigl[(v,z), (\hat v, \hat z)\bigr]\ }{(x,y)}
= 
 y_0 \cdot \contrf{v,\hat v}{\Omega^0} +
\yb_1 \cdot \contrf{v,\hat v}{\Omega^1} 
\mapob.
$$

\subheading{ An explicit example }

Let $E$ be \agvs/ of dimension $3|3$ for which we order the basis vectors
$e_1, \dots, e_6$ such that $e_1, e_2, e_3$ are even;  let
$\Omega$ be the graded skew-symmetric form given by the matrix ($j$ is the row
index)
$$
\bigl( \contrf{e_i,e_j}\Omega \bigr)_{i,j=1}^6 = 
\left(
\matrix \strut 0&1&0 &\vrule& 1&0&0       \\
         \strut -1&0&0 &\vrule& 0&0&0     \\
         \strut 0&0&0 &\vrule& 0&1&0  \vrule depth2ex width0ex height0ex   \\
\noalign{\hrule}
     \strut -1&0&0 &\vrule& 0&0&0  \vrule depth0ex width0ex height3.5ex  \\
     \strut 0&0&-1 &\vrule& 0&1&0     \\
    \strut 0&0&0  &\vrule& 0&0&-1
\endmatrix
\right)
\mapob.
$$
For the coadjoint action of $g=(a,b)$ we thus obtain
$$
\gather
x_1 \mapsto x_1 - y_0 \cdot a^2
\ ,\ \ 
x_2 \mapsto x_2 + y_0 \cdot a^1
\ ,\ \ 
x_5 \mapsto x_5 + y_0 \cdot a^5 
\ ,\ \ 
x_6 \mapsto x_6 - y_0 \cdot a^6 
\\
\xb_1 \mapsto \xb_1 - \yb_1 \cdot a^4
\ ,\ \ 
\xb_3 \mapsto \xb_3 - \yb_1 \cdot a^5
\ ,\ \ 
\xb_4 \mapsto \xb_4 + \yb_1 \cdot a^1 
\ ,\ \ 
\xb_5 \mapsto \xb_5 + \yb_1 \cdot a^3
\rlap{\mapob,}
\endgather
$$
while
all other coordinates remain unchanged. For the fundamental vector fields we
obtain
$$
\multline
(v,z)^{\Liealg g^*} =  y_0 \cdot \Bigl(
v^2 \cdot \fracp{}{x_1} -
v^1 \cdot \fracp{}{x_2} -
v^5 \cdot \fracp{}{x_5} +
v^6 \cdot \fracp{}{x_6} \Bigr)
\\
+\yb_1 \cdot \Bigl(
v^4 \cdot \fracp{}{\xb_1} +
v^5 \cdot \fracp{}{\xb_3} -
v^1 \cdot \fracp{}{\xb_4} -
v^3 \cdot \fracp{}{\xb_5} 
\Bigr)
\mapob.
\endmultline
$$
We now distinguish three cases: (i) $y_0 \neq 0$ (but real!) and $\yb_1 = 0$,
(ii) $\yb_1 \neq 0$ and $y_0 = 0$, and (iii) $y_0 \cdot \yb_1 \neq 0$. In
the first case the orbit has dimension $2|2$ with even
coordinates $x_1, x_2$ and odd coordinates $x_5, x_6$. In order to
better distinguish the even from the odd coordinates, we will change, for the
odd coordinates only, the letter $x$ to $\xi$. Substituting
the fundamental vector fields associated to basis elements $e_i$ in the
formula for the symplectic form gives us the following
identities
$$
\contrf{\fracp{}{x_2}, \fracp{}{x_1}}{\omega} = \frac{1}{y_0}
\quad,\quad
\contrf{\fracp{}{\xi_5}, \fracp{}{\xi_5}}{\omega} = \frac{1}{y_0}
\quad,\quad
\contrf{\fracp{}{\xi_6}, \fracp{}{\xi_6}}{\omega} = \frac{-1}{y_0}
\mapob,
$$
all others being either zero or determined by graded skew-symmetry.
From these identities we deduce that the even symplectic form is given as
$$
\omega = (y_0)\mo \cdot (\ dx_1 \wedge dx_2 + \tfrac12 d\xi_5 \wedge d\xi_5 -
\tfrac12 d\xi_6 \wedge d\xi_6\ )
\mapob.
$$

In the second case the orbit dimension is still $2|2$, but now with even
coordinates $\xb_4,\xb_5$ and odd coordinates $\xib_1,\xib_3$. Here we
obtain for the symplectic form the identities
$$
\contrf{\fracp{}{\xb_4}, \fracp{}{\xib_1}}{\omega} = \frac{1}{\yb_1}
\quad,\quad
\contrf{\fracp{}{\xb_5}, \fracp{}{\xib_3}}{\omega} = \frac{1}{\yb_1}
\mapob,
$$
from which we deduce that the odd symplectic form is given as
$$
\omega = (\yb_1)\mo \cdot (\ d\xib_1 \wedge d\xb_4 + d\xib_3 \wedge d\xb_5\ )
\mapob.
$$

In the third case we have to be slightly more careful. We introduce the
coordinate change $\hat x_2 = x_2$, $z_0 = y_0\, \xb_4 - \yb_1\, x_2$,
$\hat \xi_5 = \xi_5$, $z_1 = y_0\, \xib_3 + \yb_1\, \xi_5$, and then the
$z_i$ do not change under the coadjoint action. It follows that the orbit has
dimension
$3|3$ with even coordinates $x_1, \hat x_2, \xb_5$ and odd coordinates
$\xib_1, \hat \xi_5, \xi_6$. In terms of these coordinates the fundamental
vector field is given as
$$
\multline
(v,z)^{\Liealg g^*} =  y_0 \cdot \Bigl(
v^2 \cdot \fracp{}{x_1} -
v^1 \cdot \fracp{}{\hat x_2}  -
v^5 \cdot \fracp{}{\hat \xi_5}   +
v^6 \cdot \fracp{}{\xi_6} \Bigr)
\\
+\yb_1 \cdot \Bigl(
v^4 \cdot \fracp{}{\xib_1}  -
v^3 \cdot \fracp{}{\xb_5} 
\Bigr)
\mapob.
\endmultline
$$
As before, substituting suitable basis vectors for $v$ in the
formula  for the symplectic form gives us the following identities
$$
\gather
\contrf{\fracp{}{\hat x_2}, \fracp{}{x_1}}{\omega} = \frac{1}{y_0}
\quad,\quad
\contrf{\fracp{}{\hat x_2}, \fracp{}{\xib_1}}{\omega} = \frac{1}{y_0}
\quad,\quad
\contrf{\fracp{}{\xih_5}, \fracp{}{\xb_5}}{\omega} = \frac{1}{y_0}
\\
\noalign{\vskip2\jot}
\contrf{\fracp{}{\xih_5}, \fracp{}{\xih_5}}{\omega} =
\frac{1}{y_0}
\quad,\quad
\contrf{\fracp{}{\xi_6}, \fracp{}{\xi_6}}{\omega} = \frac{-1}{y_0}
\mapob.
\endgather
$$
This results in the mixed (degenerate but homogeneously non-degenerate)
symplectic form
$$
\omega = (y_0)\mo \cdot (\ {dx_1 \wedge d\hat x_2} 
+ {d\xib_1 \wedge d\hat x_2} 
+  {d\xb_5 \wedge d\hat \xi_5} 
+ \tfrac12 {d\hat \xi_5 \wedge d\hat \xi_5} 
- \tfrac12 {d\xi_6 \wedge d\xi_6}
\ )
\mapob.
$$

\heading \memorize\headnum={prequantization}. Prequantization \endheading

Let us now turn our attention to prequantization of \ssmfd/s. Since
the symplectic form can be seen as an even 2-form with values in $C$, it seems
natural that in terms of principal fiber bundles we should look at structure
groups whose associated \glalg/ is $1|1$ dimensional. A simple analysis shows
that (up to scaling) there are only three different \glalg/ structures on $C$
possible: (i) an abelian one, (ii) $\lieb{c_0,c_1} = c_1$ and $\lieb{c_1, c_1}
= 0$, and (iii) $\lieb{c_0, c_1} = 0$ and $\lieb{c_1, c_1} = c_0$. If we impose
that the even part of the group should be the circle, the second possibility
drops out because the simply connected \glgrp/ with this \glalg/ does not have
non-trivial discrete normal subgroups (it is the $a\xi + \alpha$ group of
affine transformations of the odd affine line $\CA_1$). The remaining two
possibilities both have $\Gext \SS^1 \times \CA_1$ as underlying \gmfd/. Here
$\Gext \SS^1$ is the circle augmented to even nilpotent elements, which we can
write as $\Gext \SS^1 = \{ \eexp^{ix} \mid x\in \CA_0\}$ or as $\Gext \SS^1 = \{
x
\mod 2\pi\ZZ \mid x\in \CA_0\}$. In the abelian case the group structure is
(obviously) given by 
$$
(\eexp^{ix}, \xi) \cdot (\eexp^{iy}, \eta) =
(\eexp^{i(x+y)}, \xi + \eta)
\mapob.
$$
In the non-abelian case the group structure is
given by 
$$
(\eexp^{ix}, \xi) \cdot (\eexp^{iy}, \eta) =
(\eexp^{i(x+y + \xi\eta)}, \xi + \eta)
\mapob.
$$
This group can be seen as a
$1|1$-dimensional Lie subgroup of the multiplicative subgroup of the ring $\CA
\oplus i\CA$. As such it is the $1|1$-dimensional equivalent of $\SS^1$ as the
1-dimensional Lie subgroup of the multiplicative subgroup of $\CC = \RR \oplus
i\RR$. 

If we think of prequantization in terms of complex line bundles, it is natural
to prefer the non-abelian group because it is the natural group in which
transition functions can take their values if the {line} in question is
$\CAC = \CA \oplus i\CA$. However, I could not find any reasonable (nor
unreasonable) way to disguise a (mixed) symplectic form as an \stress{even}
matrix valued 2-form such that it can be the curvature of a linear connection
on a line bundle over $M$. Since the curvature form of a linear connection is
necessarily even, this seems to exclude the non-abelian case. Remains the
abelian case, for which we have two additional arguments in favor. In the first
place: the center of the Poisson algebra is the $1|1$-dimensional abelian
subalgebra of constant functions. And secondly, the curvature 2-form of a
connection on a principal fiber bundle is a 2-form on the base space if and
only if the structure group is abelian. We thus concentrate our efforts to
answer  the question: does there exist a principal fiber bundle $\pi:Y \to M$
with abelian structure group $\Gext \SS^1 \times \CA_1$ and connection
$\double\alpha$ whose curvature is $\double\omega\,$?

To answer the above question, we follow very closely the corresponding analysis
in \cite{TW}. This involves standard techniques in algebraic topology to
show the equivalence between de~Rham cohomology and \v Cech cohomology.

\heading \recall{prequantization}$^{bis}$. Intermezzo on cohomology and \ssmfd/s
\endheading

We
start by choosing an open cover $\cover U = \{U_i \mid i\in I\}$ of the
\ssmfd/ $(M,\omega)$ such that all finite intersections of elements of $\cover U$
are contractible or empty.

\definition{\thm{Definition}}
The \stress{nerve} of the cover $\cover U$, denoted $\nerve(\cover U)$, is
defined as
$$
\nerve(\cover U) = \{ \, (i_0, \dots, i_k) \in I^{k+1} \mid k\in \NN, U_{i_0}
\cap \dots \cap U_{i_k} \neq \emptyset \,\}
\mapob,
$$
and an element $(i_0, \dots, i_k)$ is called an \stress{$k$ simplex}. The
abelian group $C_k(\cover U)$ of \stress{$k$-chains} is defined to be the free
$\ZZ$-module with basis the $k$-simplices, \ie, $C_k(\cover U)$ consists of
all finite formal sums $\sum c_{i_0, \dots, i_k}\, (i_0, \dots, i_k)$ with
$(i_0, \dots, i_k) \in \nerve(\cover U)$ and $c_{i_0, \dots, i_k} \in \ZZ$. 
The \stress{boundary operator $\partial_k : C_k(\cover U) \to C_{k-1}(\cover
U)$} is the homomorphism defined on the basis of $C_k(\cover U)$ by
$$
\partial_k(i_0, \dots, i_k) = \sum_{j=0}^k (-1)^j (i_0, \dots, i_{j-1},
i_{j+1}, \dots, i_k)
\mapob.
$$
It is an elementary exercise to prove that $\partial_{k-1} \scirc
\partial_k = 0$.

\dontprint{
\demo{Proof}
$$
\align
\partial_{k-1}\Bigl( \partial_k(i_0, &\dots, i_k) \Bigr) 
= 
\sum_{j=0}^k (-1)^j \partial_{k-1}(i_0, \dots,
i_{j-1}, i_{j+1}, \dots, i_k)
\\&
=
\sum_{j=0}^k (-1)^j \biggl(\ \sum_{\ell=0}^{j-2} (-1)^\ell (i_0, \dots,
i_{\ell-1}, i_{\ell+1}, \dots, i_{j-1}, i_{j+1}, \dots, i_k)
\\&
\hskip-2em
+(-1)^{j-1} (i_0, \dots, i_{j-2}, i_{j+1}, \dots, i_k)
+ (-1)^{j} (i_0, \dots, i_{j-1}, i_{j+2}, \dots, i_k)
\\&
\qquad
+
\sum_{\ell=j+2}^{k} (-1)^{\ell-1} (i_0, \dots, i_{j-1}, i_{j+1}, \dots,
i_{\ell-1}, i_{\ell+1}, \dots, i_k)   \ \biggr)
\\&
=
\sum_{j=0}^k \sum_{\ell=0}^{j-2} (-1)^{j+\ell} (i_0, \dots,
i_{\ell-1}, i_{\ell+1}, \dots, i_{j-1}, i_{j+1}, \dots, i_k)
\\&
\hskip-2em
- \sum_{j=1}^k (i_0, \dots, i_{j-2}, i_{j+1}, \dots, i_k)
+ \sum_{j=0}^{k-1} (i_0, \dots, i_{j-1}, i_{j+2}, \dots, i_k)
\\&
\qquad
+ \sum_{\ell=0}^{k} \sum_{j=0}^{\ell - 2} (-1)^{j+\ell-1} (i_0, \dots, i_{j-1},
i_{j+1},
\dots, i_{\ell-1}, i_{\ell+1}, \dots, i_k)
\\&
=0
\mapob,
\endalign
$$
because the two single sums in the second line cancel and because the two
double sums also cancel.
\QED\enddemo
} 

For any abelian group $A$, a homomorphism $h:C_k(\cover U) \to A$ is
completely determined by its values on the basis vectors $(i_0, \dots, i_k)
\in \nerve(\cover U)$; such a homomorphism is called a \stress{$k$-cochain} 
if it
is totally skew-symmetric on these basis vectors, \ie, $h$ changes sign when one
interchanges two entries $i_p$ and $i_q$ in $(i_0, \dots, i_k)$. The set of all
$k$-cochains with values in the abelian group $A$ is denoted by
$C^k(\cover U, A)$; equipped with pointwise addition of functions this
is an abelian group. By duality one defines the \stress{coboundary operator
$\delta_k : C^k(\cover U, A) \to C^{k+1}(\cover U, A)$}~:
$$
(\delta_k h)(i_0, \dots, i_k)
=
h(\partial_{k+1} (i_0, \dots, i_k))
\mapob.
$$
Since $\partial_{k} \scirc \partial_{k+1} = 0$, we have $\delta_{k} \scirc
\delta_{k-1} = 0$. It follows that $B^k(\cover U, A) = \im(\delta_{k-1})$
is contained in $Z^k(\cover U, A) = \ker(\delta_k)$, so their quotient
$H^k(\cover U, A) = Z^k(\cover U, A) / B^k(\cover U,
A)$ is a well defined abelian group. Elements of $B^k(\cover U, A)$ are called
\stress{$k$-coboundaries}, elements of $Z^k(\cover U, A)$ are called
\stress{$k$-cocycles}, and $H^k(\cover U, A)$ is called the \stress{$k$-th \v
Cech cohomology group of $\cover U$ with values in the abelian group $A$}.

The construction of $H^k(\cover U, A)$ can be done for any cover
$\cover U$, but it can be shown that with our restrictions (contractible
intersections) it is actually independent of the cover. It is thus customary to
denote these groups by $H^k_{\check C}(M, A)$ and to call them the $k$-th \v Cech
cohomology group of $M$ (with values in $A$).

\enddefinition

We now take a closer look at the symplectic form $\omega$.
Since the Poincar\'e lemma also
holds for \gmfd/s, there exist on each $U_i$ a 1-form $\theta_i$ such that
$d\theta_i = \omega$. On each (non empty) intersection $U_i \cap U_j$ we have
$d(\theta_i - \theta_j) = 0$ and hence there exist smooth functions $f_{ij}$ on
$U_i \cap U_j$ such that $\theta_i - \theta_j = df_{ij}$. By choosing a total order on the set
of indices $I$ we even may assume that $f_{ji} = -f_{ij}$. Hence on a triple
intersection $U_i \cap U_j \cap U_k$ we have $d(f_{ij} + f_{jk} + f_{ki}) = 0$
and hence there exist constant functions $a_{ijk}$ (necessarily real) such that
on $U_i \cap U_j \cap U_k$ we have $f_{ij} + f_{jk} + f_{ki} = a_{ijk}$. Since
the $f_{ij}$ are skew-symmetric in $i,j$, it follows immediately that the map
$a:C_2(\cover U) \to \RR$, $(i,j,k) \in \nerve(\cover U) \mapsto a_{ijk}$ is
skew symmetric, \ie, $a\in C^2(\cover U, \RR)$. Moreover, the form of the
$a_{ijk} = f_{ij} + f_{jk} + f_{ki}$ also shows that $\delta_2 a = 0$, \ie,
$a\in Z^2(\cover U, \RR)$. In other words, we have associated a 2-cocycle $a$
(with values in $\RR$) to the symplectic form $\omega$.

\proclaim{\thm{Lemma}}
The (sub)group $\Per(\omega) = \im(a: \ker \partial_2 \to \RR) \subset \RR$
depends only upon the cohomology class of $\omega$ in de~Rham cohomology, and
not upon the various choices that are possible in the construction of the
cocycle $a$.

\endproclaim

\dontprint{
\demo{Proof}
If $\theta_i$ is replaced by $\theta_i + d\phi_i$ with $\phi_i$ a smooth
function on $U_i$, then we can replace $f_{ij}$ by $f_{ij} + \phi_i - \phi_j$,
and $a_{ijk}$ remains unchanged. If $f_{ij}$ is replaced by $f_{ij} + c_{ij}$
with $c_{ij}$ a (real) constant, then $a_{ijk}$ is replaced by $(a+\delta_2
c)_{ijk}$ because $(\delta_2 c )(i,j,k) = c(\partial_2(i,j,k)) = c( (j,k) -
(i,k) + (i,j)) = c_{jk} - c_{ik} + c_{ij} = c_{ij} + c_{jk} + c_{ki}$. Since
$\delta_2 c$ is (by definition) zero on $\ker \partial_2$, this does not change
$\Per(\omega)$. This shows that $\Per(\omega)$ does not depend upon the choices
made in the construction of the cocycle $a$.  To show that it only depends upon
the cohomology class of $\omega$, we note that if we replace $\omega$ by $\omega
+ d\theta$, then we can replace $\theta_i$ by $\theta_i + \theta$, and then
$f_{ij}$ does not change.
\QED\enddemo
} 

\definition{\thm{Definition}}
The group $\Per(\omega) \subset \RR$ is called \stress{the group of periods of
the (closed) 2-form $\omega$}.

\enddefinition

\proclaim{\thmm{ainper}{Lemma}}
The
cocycle $a$ can be chosen such that $\forall (i,j,k) \in \nerve(\omega) :
a_{ijk} \in \Per(\omega)$.

\endproclaim

\dontprint{
\demo{Proof}
We define the homomorphism $b: C_1(\cover U) \to \RR/\Per(\omega)$ as follows.
On the subspace $\im \partial_2$ it is defined by  $b = \pi \scirc a \scirc
(\partial_2)\mo$, where $\pi$ denotes the canonical projection $\RR \to \RR/
\Per(\omega)$; this is independent of the choice in $(\partial_2)\mo$ by
definition of $\Per(\omega)$. Since $\RR/\Per(\omega)$is a divisible
$\ZZ$-module, there exists an extension $b$ to the whole of $C_1(\cover U)$
(see \cite{HiSt, \S1.7}). Since $C_1(\cover U)$ is a free $\ZZ$-module, there
exists a homomorphism $b' : C_1(\cover U) \to \RR$ satisfying $\pi \scirc
b' = b$. Finally we replace the functions $f_{ij}$ by $f_{ij} -
b'_{ij}$, which changes the cocycle $a$ into $a - \delta_1 b'$. By
construction of $b'$ it follows that $\pi(a_{ijk} - (\delta_1 
b')_{ijk}) = 0$, showing that this modified cocycle takes its values in
$\Per(\omega)$.
\QED\enddemo
} 

\remark{Remarks}
\itemize
The above statement is a purely algebraic statement, independent of the
topological properties of $\Per(\omega)$; the latter can be dense or discrete
in $\RR$ without affecting the result.

\itemize
The simplex $(i,j,k)$ is clearly not in $\ker \partial_2$; nevertheless the
cocycle $a$ can be chosen such that it takes everywhere values in $\Per(\omega)$.

\itemize
If $\omega$ is exact then obviously $\Per(\omega) = \nul$, but one can prove
that the converse is also true: if $\Per(\omega) = \nul$, then $\omega$ is
exact.

\itemize
The construction of the 2-cocycle $a$ associated to the closed 2-form $\omega$
is part of a larger construction which serves to prove the equivalence between
de~Rham cohomology and \v Cech cohomology. This approach can be used at the
same time to prove that the de~Rham cohomology of \agmfd/ is the same as the
de~Rham cohomology of the underlying \rmfd/ (its body).

\endremark

\dontprint{
\demo{Proof}
Suppose $\Per(\omega) = \nul$ and let $\theta_i$ and $f_{ij}$ be as in the
construction of the cocycle $a$. According to \recalt{ainper} we may assume that
the constants $a_{ijk}$ are zero, \ie, that $f_{ij} + f_{jk} + f_{ki} = 0$. Let
$\rho_i$ be a partition of unity subordinated to the cover $\cover U$. Then the
local 1-forms $\hat \theta_i =
\theta_i + d\sum_k \rho_k f_{ki}$ are well defined on $U_i$ and they coincide
on the intersection $U_i \cap U_j$~: $\hat \theta_i - \hat \theta_j = df_{ij}
 + d\sum_k \rho_k (f_{ki} - f_{kj}) = df_{ij} - d\sum_k \rho_k f_{ij} = 0$.
These local 1-forms thus define a global 1-form $\hat \theta$, which obviously
satisfies $d\hat\theta = \omega$.
\QED\enddemo
} 

\heading \recall{prequantization}. Prequantization {\rm continued}
\endheading

We now come to the construction of the principal fiber bundle $\pi : Y
\to M$ with a connection $\double\alpha$ whose curvature is $\double\omega$. 
In order to better discuss some particular details of our construction, we
choose  $d\in \RR^{\ge0}$ and we define $D = d\ZZ \subset \RR \subset \CA_0
\subset \CA$. The abelian group $\CA$ has two global coordinates $x\in \CA_0$
and $\xi\in \CA_1$. Since $D$ is a discrete subgroup of $\CA$ contained in its
even part, the quotient \glgrp/ $\CA/D = \CA_0/D \times \CA_1$ inherits a global
odd coordinate
$\xi$ and a local even coordinate $x$ modulo $d$ (of course, in case $d=0$, $x$
is a global coordinate). The sum $\cadb x = x+\xi$ of the two coordinate
functions can be seen as a (local) $\CA$-valued function. Its exterior
derivative $d\cadb x$ is a globally well defined mixed 1-form on $\CA/D$.
The abelian group $\CA/D$ is (isomorphic to) our abelian
group $\Gext\SS^1 \times
\CA_1$. The two global vector fields $\partial_x$ and $\partial_\xi$ are left
(and right) invariant; we thus can take them as a basis for the \glalg/ of
$\CA/D$. This allows us to make our choice for the \gvs/ $C$ and its basis
$c_0,c_1$ explicit:  we let $C$ be the \glalg/ of $\CA/D$ and we let
the basis $c_0,c_1$ be the left-invariant vector fields $c_0 = \partial_x$,
$c_1=\partial_\xi$. In particular the even $C$-valued 1-form $\double{d\cadb
x} = dx \otimes c_0 + d\xi
\otimes c_1$ is the Maurer-Cartan 1-form of $\CA/D$. One word of caution is in
order: the basis $c_0,c_1$ of $C$ depends upon the choice of $d\in \RR^{\ge0}$,
hence $\double\omega$ and $\double\alpha$ also depend upon this choice! Our
final assumption for the construction of the principal fiber bundle $Y$ is that 
$\Per(\omega)$ is contained in $D$. 

And then the actual construction. Since we assume that $a_{ijk}\in \Per(\omega)$,
it follows that the functions
$$
g_{ij} : U_i \cap U_j \to \CA /D
\quad,\quad
m\mapsto f_{ij}(m) \mod D
$$
satisfy the cocycle condition
$$
g_{ij} + g_{jk} + g_{ki} = 0 \qquad\text{on $U_i\cap U_j \cap U_k$.}
\formula{cocyclecondition}
$$
Hence these functions define a principal fiber bundle $\pi:Y \to M$ 
with structure group $\CA /D$. Its local trivializations are $U_i \times \CA
/D$ with projection $\pi(m,\cadb x_i) = m$ and transition functions 
$$
\aligned
U_i \times \CA /D
&\to
U_j \times \CA /D
\\
(m,\cadb x_i) &\mapsto 
(m,\cadb x_i+ g_{ij}(m)) = (m,\cadb x_j)
\mapob.
\endaligned
\formula{Ytrans}
$$
On $Y$ we define the 1-form $\alpha$ by its expression on each local chart $U_i
\times \CA/D$ by
$$
\alpha = \pi^*\theta_i + d\cadb x_i
\formula{locexpralpha}
$$
satisfying $d\alpha = \pi^*\omega$. That $\alpha$ is well defined
follows from the definition of the functions $g_{ij}$ and the fact that 
$dg_{ij} = df_{ij}$ (same argument as in proving that $d\cadb x$ is a global
1-form on $\CA/D$). With our choice for
$c_0,c_1$ it is elementary to verify that 
$$
\double\alpha \equiv \bigl( (\theta_i)_0 + dx_i\bigr) \otimes c_0 + \bigl(
(\theta_i)_1 + d\xi_i \bigr) \otimes c_1 
$$ 
is a
connection 1-form on $Y$ whose curvature 2-form is
$\double\omega$.

\dontprint{
\demo{Proof}
Since the right action of $\CA /D$ on $Y$ is given on a local chart
by
$(m,\cadb x_i) \cdot  \cadb t = (m,\cadb x_i+\cadb t)$, it follows
immediately that this action preserves the 1-form $\alpha$ and thus
$\double\alpha$. Since the group is abelian, the adjoint action is trivial and
thus the second condition of a connection 1-form is satisfied. For the first
condition, we note that the fundamental vector fields associated to the Lie
algebra elements $c_0 = \partial_{t}, c_1 =
\partial_{\tau}$ are $\partial_{x_i}$ and 
$\partial_{\xi_i}$. It follows that the
contraction of $\double\alpha$ with a fundamental vector field gives the
initial Lie algebra element. This proves that $\double\alpha$ is a connection
1-form. That $\double\omega$ is its curvature is (again) a direct consequence
of the fact that the structure group is abelian: for abelian structure groups
the curvature 2-form is the exterior derivative of the connection 1-form.
\QED\enddemo
}  

\proclaim{\thmm{Ybyconstruction}{Theorem}}
There exists a principal fiber bundle $\pi: Y \to M$ with structure group $\CA
/D$ and connection $\double\alpha$ whose curvature is $\double\omega$ if and only
if $\Per(\omega)$ is contained in $D$, in which case $Y$ is obtained by the
above construction.

\endproclaim

\dontprint{
\demo{Proof}
We only have to prove the only if part, so suppose we have such a bundle $Y$.
Without loss of generality we may assume that $Y$ is trivial above each
$U_i$. We thus have transition functions $g_{ij} : U_i \cap U_j \to \CA/D$. 
Since $\CA \to \CA/D$ is a covering and since each $U_i \in \cover U$ is
contractible, there exist smooth functions $f_{ij} : U_i \cap U_j \to \CA$ such
that $g_{ij} = f_{ij} \mod D$. On the trivializing chart $U_i \times \CA/D$ the
connection $\double\alpha$ is necessarily of the form
$$
\double\alpha \equiv \bigl( (\theta_i)_0 + dx_i\bigr) \otimes c_0 + \bigl(
(\theta_i)_1 + d\xi_i \bigr) \otimes c_1 = 
\double{\theta_i + d\cadb x_i}
$$ 
for some local 1-form $\theta_i$. The fact that the curvature of
$\double\alpha$ is $\double\omega$ implies that $d\theta_i = \omega$. Comparing
the local expressions above $U_i$ and $U_j$ for the global connection
$\double\alpha$ using the transition \recalf{Ytrans} gives us
$$
\theta_j + d(\cadb x_i + g_{ij}) = \theta_i + d\cadb x_i
\mapob.
$$
Hence $\theta_i - \theta_j =dg_{ij} \equiv df_{ij}$. It now
suffices to note that the functions $g_{ij}$ satisfy the cocycle condition
\recalf{cocyclecondition} to conclude that $a_{ijk} \in D$. And thus
$\Per(\omega) \subset D$. From the above analysis the last statement is also
immediate.
\QED\enddemo
}  

Instead of performing our construction of $Y$ with the $\CA$-valued functions
$f_{ij}$, we could have restricted our attention to the even part only. This
means that we use the functions $(g_{ij})_0 : U_i \cap U_j \to \CA_0/D$ to
define a principal fiber bundle $\YO \to M$ with structure group $\CA_0/D$.
Another way to describe $\YO$ is to say that it is the subbundle of $Y$
corresponding to the subgroup $\CA_0/D \subset \CA/D$. Said this way, it is
clear that the construction of $\YO$ is intrinsic. We
also only consider the even part of $\alpha$~: $\alpha_0 =
(\pi^*\theta_i)_0 + dx_i$. And as for $\alpha$, $\alpha_0 \otimes c_0$ is a
connection 1-form on the principal fiber bundle $\YO$. We could and will say
that \stress{$(\YO,\alpha_0)$ is the even part of $(Y,\alpha)$}. The importance
of the even part of $Y$ lies in the result [8.10], which needs  a
definition.

\definition{\thm{Definitions}}
A principal fiber bundle $\pi : Y \to M$ with structure  group $G$ is called
\stress{topologically trivial} if there exists a global smooth section $s:M \to
Y$. A \stress{$D$-prequantum bundle for the symplectic manifold $(M,\omega)$}
is a principal fiber bundle $\pi : Y \to M$ with structure group $\CA/D$ and
connection $\double\alpha$ whose curvature is $\double\omega$.  A $D$-prequantum
bundle will usually be denoted as a couple $(Y,\alpha)$. Two
$D$-prequantum bundles $(Y,\alpha)$ and $(Y',\alpha')$ are called
\stress{equivalent (as $D$-prequantum bundles)} if there exists a diffeomorphism
$\phi:Y \to Y'$ such that $\pi' \scirc \phi = \pi$, commuting with the
 right-actions of $\CA/D$ in the sense that $\phi(y\cdot \cadb t) = \phi(y)
\cdot \cadb t$, and such that $\phi^*\alpha' = \alpha$.

\enddefinition

\proclaim{\thmm{triviality}{Theorem}}
Let $(Y,\alpha)$ be a $D$-prequantum bundle.
\roster
\item"(i)"
If $D=\nul$, then $Y$  is topologically trivial.

\item"(ii)"
If $Y$ is a topologically trivial, then $\Per(\omega) =
\nul$ and $(Y,\alpha)$ is equivalent to the
$D$-prequantum bundle
$M\times \CA/D$ with connection $\double{\theta + d\cadb x}$, where $\theta$ is
a global 1-form satisfying $d\theta = \omega$.

\item"(iii)"
$Y$ is equivalent to the $D$-prequantum bundle $\YO \times \CA_1$ with
connection ${\alpha_0 \otimes c_0} + (\theta_1 + d\xi) \otimes c_1$, where
$\theta_1$ is a global odd 1-form on $M$ satisfying $d\theta_1 = \omega_1$.

\item"(iv)"
Inequivalent $D$-prequantum bundles are classified by
$H^1_{\check C}(M, \RR/D)$.

\endroster

\endproclaim

\dontprint{
\demo{Proof}
According to \recalt{Ybyconstruction} we assume that $Y$ is constructed with the
ingredients $\theta_i$, $f_{ij}$ and $a_{ijk} \in \Per(\omega)$ associated to the
cover $\cover U$. We let $\rho_k$ be a partition of unity subordinated to the
cover $\cover U$.

\itemize
(i)
If $D=\nul$, then $\Per(\omega) = \nul$ because it is supposed to be included
in $D$, but also the functions $f_{ij}$ act as transition functions for the
principal fiber bundle.  We define local sections $s_i : U_i \to \pi\mo(U_i)
\cong U_i \times \CA$ in terms of the local trivializations by
$$
s_i(m) = (m, \sum_k \rho_k(m) f_{ki}(m)\,)
\mapob.
$$
In order to show that these local sections glue together to a global section we
first note that since $\Per(\omega) = \nul$, all constants $a_{ijk}$ are zero.
This together with the skew-symmetry of the $f_{ij}$ implies that $f_{ij} +
f_{jk} = f_{ik}$. Then we recall that the transition functions of $Y$ are given
by \recalf{Ytrans}, \ie, $s_i(m)$ is mapped to $(m,f_{ij}(m) + 
\sum_k \rho_k(m) f_{ki}(m)\,)$. But 
$$
f_{ij}(m) + 
\sum_k \rho_k(m) f_{ki}(m) = \sum_k \rho_k(m) (f_{ki}(m) + f_{ij}(m))
=
\sum_k \rho_k(m) f_{kj}(m)
\mapob.
$$
In other words, $s_i(m)$ is mapped by the transition functions to $s_j(m)$.
Hence the local sections $s_i$ glue together to form a global smooth section.

\itemize
(ii)
If $s:M\to Y$ is a global section, we can define the 1-form $\theta =
s^*\alpha$ on $M$. Obviously $d\theta = s^*d\alpha = s^*\pi^*\omega = \omega$.
Hence $\omega$ is exact and thus $\Per(\omega) = \nul$. We now define the map
$\phi : M \times \CA/D \to Y$ by
$$
\phi(m,\cadb x) = s(m) \cdot \cadb x
\mapob,
$$
where on the right hand side we use the action of the structure group $\CA/D$
on $Y$. If the section $s$ is represented on a local trivializing chart $U_i$
by the function $s_i:U_i \to \CA/D$, $s(m) = (m,s_i(m))$, then $\phi$ is given by
$\phi(m,\cadb x) = (m,s_i(m) + \cadb x)$, from which it follows immediately that
$\phi$ is a diffeomorphism such that $\pi \scirc \phi = \pi_1$ ($\pi_1:M\times
\CA/D
\to M$ the canonical projection) and commuting with the $\CA/D$-action. 
In the local trivialization we also have $\alpha = \theta_i + d\cadb x_i$,
and thus, since $s^*\alpha = \theta$, we have $\theta = \theta_i + ds_i$ on
$U_i$. Finally, still in the same trivializing chart we have
$\phi^*\alpha = \theta_i + d(s_i + \cadb x) = \theta + d\cadb x$.

\itemize
(iii)
Since the constants
$a_{ijk}$ are real, we have $(f_{ij})_1 = (g_{ij})_1$. We define the
isomorphism $\phi: Y \to \YO \times \CA_1$ on local trivializing charts $U_i
\times (\CA_0/D) \times \CA_1$ by
$$
\phi(m,\cadb x_i) = (m,\cadb x_i + \sum_k \rho_k(m) (f_{ik}(m))_1 )
\mapob.
$$
It is obvious that this is a diffeomorphism between the local trivializing
charts, commuting with the $\CA/D$ action and compatible with the bundle
structure. Remains to be verified that it is globally well defined. For the
bundle $Y$ the transition functions are given by \recalf{Ytrans}, for
$\YO\times \CA_1$ they are given by $(m,x_i+\xi) \mapsto (m, x_i +
(g_{ij}(m))_0 + \xi)$. We now start with a point $(m,\cadb x_i) \in U_i
\times \CA/D \subset Y$. If we first apply $\phi$ and then change charts to
$U_j \times \CA/D$ in $\YO\times \CA_1$, we obtain 
$$
(m, \cadb x_i + (g_{ij}(m))_0 + \sum_k \rho_k(m) (f_{ik}(m))_1 )
\mapob.
$$ 
On the
other hand, if we first change charts to $U_j \times \CA/D$ in $Y$ and then
apply $\phi$, we obtain 
$$
(m, \cadb x_i + (g_{ij}(m))_0 + (f_{ij}(m))_1 + \sum_k \rho_k(m)
(f_{jk}(m))_1 ))
\mapob.
$$ 
Since $(f_{ij}(m))_1 + (f_{jk}(m))_1 =-  (f_{ki}(m))_1 =  (f_{ik}(m))_1$
according to \recalf{cocyclecondition}, these two results are the same. Hence
$\phi$ is a well defined global isomorphism of principal fiber bundles.

Under $\phi\mo$ the local odd 1-forms $(\theta_i)_1$ change to $\thetah_i =
(\theta_i)_1 - d\sum_k \rho_k (f_{ik})_1$. On non-empty
intersections $U_i \cap U_j$ we
have $\thetah_i - \thetah_j = d ( f_{ij} - \sum_k \rho_k( f_{ik} -
f_{jk}) \,)_1 = {d(f_{ij} - \sum_k \rho_k f_{ij})_1} = 0$. It follows that the
local odd 1-forms $\thetah_i$ glue together to form a global odd 1-form
$\theta_1$, which obviously satisfies $d\theta_1 = \omega_1$. By construction we
have $(\phi\mo)^* \alpha = \alpha_0 + \theta_1 + d\xi$, which finishes the proof
of (iii).

\itemize
(iv)
Let $(Y^r,\alpha^r)$ be a fixed (reference) $D$-prequantum bundle constructed
with the ingredients $\theta^r_i$, $f^r_{ij}$, $a^r_{ijk}\in \Per(\omega)$ and
let $(Y,\alpha)$ be an arbitrary $D$-prequantum bundle constructed
with the ingredients $\theta_i$, $f_{ij}$, $a_{ijk}\in \Per(\omega)$. Since
$U_i$ is contractible, there exist $g_i:U_i \to \CA$ such that $\theta_i -
\theta_i^r = dg_i$. Since $U_i\cap U_j$ is contractible, there exist constants 
$b_{ij} \in \RR$ such that $f_{ij} - f^r_{ij} = g_i - g_j + b_{ij}$. Since
$a_{ijk},a^r_{ijk} \in \Per(\omega) \subset D$, it follows that $\pi b_{ij} + \pi
b_{jk} + \pi b_{ki} = 0$, where $\pi$ denotes the projection $\pi: \RR \to
\RR/D$ (abuse of notation without confusion). Hence the 1-cochain $\pi b_{ij}$
is actually a 1-cocycle and thus determines an element $[\pi b] \in
H^1_{\check C}(M,\RR/D)$. The only freedom in the construction of this cohomology
class is the choice of $g_i$~; if we change $g_i$ to $g_i + c_i$ for constants
$c_i \in \RR$, the cocycle $\pi b$ is changed to $\pi b + \pi \delta_1 c$, and
thus the cohomology class $[\pi b]$ is independent of this choice. We thus have
constructed a map from the set of $D$-prequantum bundles $(Y,\alpha)$ to
$H^1_{\check C}(M,\RR/D)$.  The next steps are to prove that this induces a
bijection on equivalence classes of $D$-prequantum bundles.

Let us first assume that $(Y,\alpha)$ and $(Y',\alpha')$ determine the same
cohomology class, \ie, there exist constants $c_i \in \RR$ such that
$\pi b_{ij}' = \pi b_{ij} + \pi c_j - \pi c_i$. From the definitions it follows
that we have
$$
\pi f'_{ij} - \pi f_{ij} = \pi(g'_i -g_i + c_i) - \pi(g'_j - g_j + c_j)
\mapob.
\formula{YY'samecoh}
$$
We now define a map $\phi: Y \to Y'$ on local trivializing charts by
$\phi(m,\cadb x_i) = (m,\cadb x_i - \pi(g'_i(m) -g_i(m) + c_i))$. If we
first change charts in $Y$ to $U_j$ and then apply this $\phi$ we obtain
$$
(m, \cadb x_i + \pi f_{ij}(m)  - \pi(g'_j(m) -g_j(m) + c_j))
\mapob,
$$
while if we first apply this $\phi$ and then change charts in $Y'$ we obtain
$$
(m, \cadb x_i - \pi(g'_i(m) -g_i(m) + c_i) + \pi f'_{ij}(m))
\mapob.
$$
According to \recalf{YY'samecoh} these two results are the same, showing that
$\phi$ is a globally well defined diffeomorphism satisfying $\pi' \scirc \phi =
\pi$ and compatible with the action of $\CA/D$. Since $\theta'_i - \theta_i =
dg'_i - dg_i$, it follows that $\phi^*(\theta'_i + d\cadb x_i) = \theta'_i +
d\cadb x_i - (dg'_i - dg_i) = \theta_i + d\cadb x_i$. In other words,
$\phi^*\alpha' = \alpha$. This shows that $D$-prequantum bundles mapping to the
same cohomology class are equivalent.

Conversely, suppose that $(Y,\alpha)$ and $(Y',\alpha')$ are equivalent via
the diffeomorphism $\phi:Y \to Y'$. This implies that there exist smooth
functions $\chi_i : U_i \to \CA/D$ such that on a local trivializing chart we
have
$$
\phi(m,\cadb x_i) = (m, \cadb x_i + \chi_i(m))
\mapob.
$$
Since $U_i$ is contractible, there exist smooth functions $h_i : U_i \to \CA$
such that $\pi h_i = \chi_i$. The fact that $\phi$ is globally defined implies
that these functions $h_i$ satisfy the condition $\pi(h_i + f'_{ij}) =
\pi(f_{ij} + h_j)$. The condition that $\phi^*\alpha' = \alpha$ translates to
the fact that $\phi^*(\theta'_i + d\cadb x_i) = \theta_i + d\cadb x_i$.
This gives us $\theta'_i -\theta_i = -dh_i$. Since this is also equal to $dg_i'
- dg_i$, there exist constants $c_i$ such that $g'_i - g_i + h_i = c_i$ ($U_i$
is connected). If we now apply the definitions, we obtain without difficulty
that $\pi b'_{ij} - \pi b_{ij} = \pi c_i - \pi c_j$, \ie, $\pi b' - \pi b =
\pi \delta_0 c$ and thus $Y$ and $Y'$ determine the same cohomology class. We
conclude that the map from equivalence classes of $D$-prequantum bundles to
$H^1_{\check C}(M,\RR/D)$ is injective.

To finish the proof, let $b_{ij}$ be constants such that $\pi b$ is a
1-cocycle. We have to construct a $D$-prequantum bundle $(Y,\alpha)$ which
determines this cocycle under the map from $D$-prequantum bundles to
$H^1_{\check C}(M,\RR/D)$. The reference bundle is constructed with the
transition functions $\pi f^r_{ij}$. Since $\pi b$ is a cocycle, the functions
$\pi f_{ij}$ with $f_{ij} = f^r_{ij} + b_{ij}$ also satisfy the cocycle
condition \recalf{cocyclecondition}. Since the $b_{ij}$ are constants, the local
1-forms $\theta^r_i + d\cadb x_i$ still glue together to form a global
1-form (connection) $\alpha$. We thus obtain a $D$-prequantum bundle
$(Y,\alpha)$, and it is immediate that this bundle maps to the cohomology class
$[\pi b]$.
\QED\enddemo
} 

\remark{Remarks}
\itemize
If $D=\nul$, then necessarily $\Per(\omega) = \nul$ (because it is contained in
$D$) and thus part (ii) of \recalt{triviality} is a partial converse to part
(i). However, it is possible that $\Per(\omega) = \nul$ and that $Y$ is not a
topologically trivial bundle, but this can happen only if $D$ is different from
$\nul$ (see \cite{TW} for an explicit example).

\itemize
Part (iii) of \recalt{triviality} can be interpreted in different ways. In the
first place, $\CA_1$ is a vector space and thus a simple partition of unity
argument shows that any principal fiber bundle with structure group $\CA_1$ is
topologically trivial (see \cite{Hi}). In the second place, any closed
\stress{odd} 2-form is exact, and thus, if we perform our construction with a
single $U=M$, we directly obtain the direct product $M\times \CA_1$ with
connection $\theta_1 + d\eta$ for the odd-part of the principal bundle.

\itemize
It is a standard result in algebraic topology that $H^1_{\check C}(M,\RR/D)$ is
isomorphic to $\Hom(\pi_1(M) \to \RR/D)$. With this result we can give another
interpretation of \recalt{triviality}(iv). If $(Y^r, \alpha^r)$ is a reference
bundle, we denote by $(\overline Y^r, \overline\alpha^r)$ the $D$-prequantum
bundle over the simply connected covering $\overline M$ of $M$ obtained by
pull-back. It follows that $\pi_1(M)$ acts on $\overline Y^r$ commuting with
the right action of $\CA/D$, and that $Y^r =
\overline Y^r/\pi_1(M)$. Now the various inequivalent $D$-prequantum bundles
can be obtained by the following procedure. For any homomorphism $\phi:
\pi_1(M) \to \RR/D$ we define a modified action $\Phi_\phi:\pi_1(M) \times
\overline Y^r \to \overline Y^r$ of $\pi_1(M)$ on $\overline Y^r$ by 
$$
\Phi_\phi(g,\overline y) = \Phi(g,\overline y) \cdot \phi(g\mo)
\mapob,
$$
where $\Phi$ denotes the previously mentioned action of $\pi_1(M)$ on $\overline
Y^r$. This is a (left) action because $\phi$ is a homomorphism and because
$\Phi$ commutes with the $\CA/D$ action. Taking the quotient of $(\overline
Y^r, \overline\alpha^r)$ with respect to this modified action gives us a
$D$-prequantum bundle $(Y_\phi, \alpha_\phi)$. Varying the homomorphism $\phi$
gives all inequivalent $D$-prequantum bundles.

\endremark

\bigskip

Readers might have wondered why we introduced the subgroup $D\subset \RR$. The
reason is simply to have a natural way to state the prequantization
construction according to Souriau. According to Souriau, a prequantization of a
symplectic manifold $(M,\omega)$ is a $D$-prequantum bundle $(Y,\alpha)$ with
$D = 2\pi\hbar \ZZ$. This is always possible if $\Per(\omega) = \nul$, it gives
a quantization condition if $\Per(\omega)$ is discrete (if $\Per(\omega) =
\lambda\ZZ$, the condition is $\lambda\in 2\pi\hbar\ZZ$), and it is never
possible if $\Per(\omega)$ is dense in $\RR$.

After this discussion on $D$-prequantum bundles leading to prequantization
in the sense of Souriau, we now come back to prequantization in the sense of
Kostant. This means looking for a complex line bundle $L$
over $M$ with connection $\nabla$ whose curvature is $i\omega/\hbar$.
Unfortunately, I am unable to define a vector bundle over $M$ with
typical fiber the ``complex line'' $\CAC = \CA \oplus i\CA = \CA \otimes_\RR
\CC$ and linear connection $\nabla$ whose curvature is the (mixed) symplectic
form. The reason is that the curvature of a linear connection is necessarily
even. 
However, there is an answer if we are slightly less demanding. A reasonable way
to obtain a connection on a vector bundle is to start with a principle fiber
bundle with a connection and to construct an associated vector bundle via a
representation of the structure group. This is what is done in the ungraded
case and which provides there the equivalence between the approaches of Souriau
and Kostant. In the graded case a natural representation $\rho$ of $\CA/D$ on
$\CAC$ is given by $\rho(t,\tau) =
\eexp^{2\pi it/d}$, where $\eexp^{iz}$ should be interpreted as the (even)
automorphism of
$\CAC$ of multiplication by
$\eexp^{iz}$. This representation is injective on $\CA_0/D$ but ignores the odd
part of $\CA/D$. When one computes the curvature of the connection $\nabla$
induced on the associated vector bundle by the connection $\double\alpha$ on
$Y$, one finds
$$
\curv(\nabla) = \rho_* \curv(\double\alpha) =  \frac{2\pi i}{d} \cdot id(\CAC)
\cdot
\omega_0
\mapob.
$$
As was to be expected, we only recover the even part of $\omega$ in the
curvature of the connection $\nabla$. Note that the \gvs/ $C$ is not involved:
the curvature of $\nabla$ is a 2-form with values in the endomorphisms of
$\CAC$, which is canonically isomorphic to $\CAC$ (matrices of size $1\times
1$). It also follows immediately that if we
want this curvature to be (the even part of) $i\omega/\hbar$, we have to choose
$d=2\pi\hbar$, which gives in turn, via $\Per(\omega) \subset d\ZZ$, the
quantization condition that a generator of $\Per(\omega)$ should be a multiple
of $2\pi\hbar$. These heuristic arguments can be made rigorous and give the
following proposition.

\proclaim{\thm{Proposition}}
There exists an $\CAC$-line bundle $L$ over $M$ with connection $\nabla$ whose
curvature is $i \omega_0/\hbar$ if and only if $\Per(\omega) \subset 2\pi\hbar
\ZZ$. If that is the case, $L$ is the line bundle associated to a
$D$-prequantum bundle $Y$ with $D\equiv 2\pi\hbar \ZZ$ by the representation
$\rho(x,\xi) =  \eexp^{ix/\hbar}$ of
$D$ on $\CAC$, .

\endproclaim

\dontprint{
\demo{Proof}
\itemize
Suppose first that $\Per(\omega) \subset D\equiv 2\pi\hbar\ZZ$. Then there
exists a 
$D$-prequantum bundle $Y$. Using the representation $\rho:
D \to \End(\CAC)$, $\rho(x,\xi) = \eexp^{ix/\hbar}$, we then form
the associated line bundle $L$ with the connection $\nabla$ induced from the
connection $\double\alpha$ on $Y$. The curvature of this connection is given by
the formula
$$
\curv(\nabla) = \rho_*\curv(\double\alpha) = i \omega_0/\hbar
\mapob,
$$
where we used $\rho_* \partial_x = i/\hbar$ and $\rho_*\partial_\xi = 0$.
This proves the if part and the second statement.

\itemize
Next we suppose that $(L,\nabla)$ exists and we use a cover $\cover U$ as in
\S\recall{prequantization}$^{bis}$ such that $L$ is trivial above each $U_i$.
The standard (partition of unity) argument that the structure group of a vector
bundle can be reduced to the orthogonal group applies here as well and shows
that we may assume that the (even) transition functions $g_{ij} : U_i \cap U_j
\to
\Aut(\CAC) \cong \{x\in \CA^{\CC}_0 \mid \body x \neq 0\}$ are of the form
$g_{ij}(m) = \eexp^{i\varphi_{ij}(m)}$ for some function $\varphi_{ij} : U_i \cap U_j
\to \CA_0$. Since we assume that $U_i \cap U_j$ is contractible, we also may
assume that $\varphi_{ij}$ is smooth. 

If $s$ is a (global) section of $L$, it is locally above $U_i$ represented by a
function $s_i:U_i \to \CAC$. If $X$ is a global vector field, the covariant
derivative $\nabla_X s$ is locally represented by the function $(\nabla_X s)_i$
given by the expression
$$
(\nabla_X s)_i = Xs_i + \contrf{X}{\Gamma_i} s_i
$$
for some even $\CAC$-valued 1-form $\Gamma_i$ on $U_i$. The curvature of the
connection $\nabla$ is locally given by the $\CAC$-valued 2-form $d\Gamma_i$
(the group $\Aut(\CAC)$ is abelian, so the term
$\tfrac12\lieb{\Gamma_i,\Gamma_i}$ in $D\Gamma_i = d\Gamma_i +
\tfrac12\lieb{\Gamma_i,\Gamma_i}$ vanishes). On the overlap of two trivializing
charts $U_i$ and $U_j$ the local functions $s_i$ and $s_j$ are related by $s_j
= s_i \cdot g_{ij}$ and similarly $(\nabla_X s)_j = (\nabla_X s)_i \cdot
g_{ij}$. This gives us the relation
$$
\Gamma_i = id\varphi_{ij} + \Gamma_j
\mapob,
$$
because $g_{ij} =
\eexp^{i\varphi_{ij}}$ is even and thus commutes with everything.
Choosing a global potential $\theta_1$ for $\omega_1$, \ie, $d\theta_1 =
\omega_1$, we introduce the 1-forms $\theta_i = -i\hbar \Gamma_i + \theta_1$. We
also introduce and the functions
$f_{ij} = \hbar\varphi_{ij}$. With these definitions we have on the one hand
$d\theta_i = \omega$ (the curvature of $\nabla$ is
$i\omega_0/\hbar$) and on the other hand $\theta_i - \theta_j = df_{ij}$.
Since the transition functions $g_{ij}$ satisfy the cocycle condition
\recalf{cocyclecondition} (in multiplicative notation), it follows that
$\varphi_{ij} + \varphi_{jk} + \varphi_{ki} \in 2\pi\ZZ$, and thus the
constants $a_{ijk}$ constructed in \S\recall{prequantization}$^{bis}$ are in
$2\pi\hbar\ZZ \equiv D$. This implies directly that $\Per(\omega) \subset
D$ (we have seen that the odd part of $\omega$ does not contribute
to the group of periods). And thus we have proved the only if part. Moreover,
it is immediate from the above construction that we have a
$D$-prequantum bundle whose transition functions $f_{ij} \mod
D$ map under the representation $\rho$ exactly to the transition
functions $g_{ij}$ of $L$. Hence $L$ can be seen as the $\CAC$-line bundle
associated to this prequantum bundle.
\QED\enddemo
} 

We see that the weakened prequantization in the sense of Kostant exists under
exactly the same conditions as prequantization in the sense of Souriau and that
both are directly related.
Some of the motivations that led to the introduction of prequantization are
given in the next section, as well as an argument that it is reasonable to
weaken prequantization in the sense of Kostant for mixed symplectic forms.

\remark{Remark}
Most texts on prequantization
in terms of line bundles (e.g., \cite{Ko1}, \cite{Sn}, \cite{Wo}) also talk
about compatible inner products, something which has been ignored in this
discussion. This additional structure has two purposes. It is used in the
definition of equivalent line bundle prequantizations, and it is used in the
definition of a scalar product on a functions space. We skipped the first item
because it will give us exactly the same classification as for $D$-prequantum
bundles
\recalt{triviality}. And the second item is beyond the scope of this paper.

\endremark

\heading \headnum. Representations and invariant subspaces \endheading

Quantization and representation theory both are interested in (a kind of)
irreducible representations. We start this section by investigating in more
detail the representation aspect of prequantization in the sense of Souriau.

\definition{\thm{Definition}}
An \stress{infinitesimal symmetry of a $D$-prequantum bundle $(Y,\alpha)$} is a
(smooth) vector field $Z$ on $Y$ preserving the connection $\double\alpha$, \ie,
the Lie derivative of $\double\alpha$ in the direction of $Z$ is zero:
$\Lied(Z)\double\alpha = 0$. This is equivalent to the condition that $Z$
preserves the homogeneous parts of $\alpha$~: $\Lied(Z)\alpha_0 = 0 = \Lied(Z)
\alpha_1$. The set of all infinitesimal symmetries of $(Y,\alpha)$ is denoted
by $\Sym(Y,\alpha)$~; it is a subset of the set of all (smooth) vector fields
on $Y$.

\enddefinition

\proclaim{\thmm{infsymprop}{Proposition}}
Infinitesimal symmetries enjoy the following properties.

\roster
\item"(i)"
$\Sym(Y,\alpha)$ is a Lie algebra when equipped with the
commutator of vector fields.

\item"(ii)"
For each $f\in \Poisson$ there exists a unique $\eta_f \in \Sym(Y,\alpha)$ such
that $\contrf{\eta_f}{\double\alpha} =- \pi^* f$.

\item"(iii)" The map $f\mapsto \eta_f$, $\Poisson \to \Sym(Y,\alpha)$ is an
isomorphism of \rglalg/s with the additional property that $\pi_* \eta_f = X_f$.

\item"(iv)"
In a local chart $U_i \times
\CA/D$ the vector field $\eta_f$ takes the form
$$
\eta_f = X_f - \Bigl( f^0 + \contrf{X_f}{(\theta_i)_0} \Bigr) \fracp{}{x_i}
- \Bigl( f^1 + \contrf{X_f}{(\theta_i)_1} \Bigr) \fracp{}{\xi_i}
\mapob.
\formula{loxexpreta}
$$

\endroster

\endproclaim

\dontprint{
\demo{Proof}
\itemize
(i): 
Since $\double\alpha$ is even, if $Z$ preserves $\double\alpha$, its
homogeneous parts $Z_0,Z_1$ do too. In order to show that the commutator of two
infinitesimal symmetries is again an infinitesimal symmetry, we thus may assume
that they are homogeneous. But then the result is obvious because for
homogeneous vector fields $Z$ and $Z'$ we have $\lieb {\Lied(Z), \Lied(Z')} =
\Lied(Z) \scirc \Lied(Z') \pm \Lied(Z') \scirc \Lied(Z)$ with the sign
determined by the parities. 

\itemize
(ii)--(iv):
Let $Z\in \Sym(Y,\alpha)$ be arbitrary. From the equation $\Lied(Z)
\double\alpha = 0$ and the fact that $d \double\alpha = \pi^* \double \omega$ we
deduce that at each point $y\in Y$ we have
$$
(d \contrf{Z}{\double\alpha})\restricted_y + \pi^* \contrf{\pi_*
Z\restricted_y}{\double\omega\restricted_{\pi(y)}} = 0
\mapob.
$$
This implies that the derivatives of the function $\contrf{Z}{\double\alpha}$
in the direction of the fiber coordinates must be zero. Since the fibers are
connected this implies that this function is independent of the fiber
coordinates, and thus that there exists a (unique) function $f\in \Ci(M,C)$
such that $\contrf{Z}{\double\alpha} = -\pi^* f$. But then, using that $\pi^*$
is injective, we have the equation
$$
\contrf{\pi_* Z\restricted_y}{\double\omega\restricted_{\pi(y)}}
-df\restricted_{\pi(y)} = 0
\mapob.
$$
This implies that $\pi_*Z\restricted_y$ depends only upon $\pi(y)$ ($\omega$
is homogeneously non-degene\-rate), \ie, 
$\pi_*Z$ is a well defined vector field on $M$. But then we have the global
equation $\contrf{\pi_* Z}{\double\omega}
=df$, which shows that $f$ belongs to $\Poisson$ and that $\pi_*Z = X_f$ is the
associated hamiltonian vector field.

Now let $Z'\in\Sym(Y,\alpha)$ be another infinitesimal symmetry, with
$\contrf{Z'}{\double\alpha} = -\pi^* df'$, $f'\in \Poisson$. We then compute:
$$
\align
\contrf{\lieb{Z,Z'}}{\double\alpha}
&=
\lieb{\Lied(Z), \contrf{Z'}{}} \double\alpha
=
\Lied(Z) \contrf{Z'}{\double\alpha} - \sum_{\beta,\gamma = 0}^1
\comm\beta\gamma \contrf{Z'_\beta}{\Lied(Z_\gamma)\double\alpha}
\\
&=
-\pi^*\Lied(\pi_*Z)f'
=
-\pi^* X_f f' = -\pi^*\PB{f,f'}
\mapob.
\endalign
$$
This shows that the map $Z\mapsto f$ from $\Sym(Y,\alpha)$ to $\Poisson$ is a
morphism of \glalg/s. Since $\double\alpha$ is even, this map is also even. It
thus remains to prove that it is bijective. For $f\in \Poisson$ we thus have to
find $Z\in \Sym(Y,\alpha)$ such that $\contrf{Z}{\double\alpha} = -\pi^*df$,
which implies that $\pi_* Z = X_f$. In a local trivializing chart $U_i \times
\CA/D$ the looked for vector field $Z$ thus must be of the form
$$
Z\restricted_{(m,\cadb x_i)} = X_f\restricted_m + g(m,\cadb x_i)
\fracp{}{x_i}\restricted_{(m,\cadb x_i)} + \chi(m,\cadb x_i)
\fracp{}{\xi_i}\restricted_{(m,\cadb x_i)}
\mapob,
$$
for some local functions $g$ and $\chi$. Since $\double\alpha$ has the local
form
$\double\alpha = \double{\theta_i + d\cadb x_i}$ \recalf{locexpralpha}, the
condition
$\contrf{Z}{\double\alpha} = -\pi^*df$ gives us 
$$
g = -f^0 - \contrf{X_f}{(\theta_i)_0} \quad,\quad
\chi = -f^1 - \contrf{X_f}{(\theta_i)_1}
\mapob.
$$
It follows that $Z$ is uniquely determined on the local trivializing  chart $U_i
\times \CA/D$ by the equations $\pi_*Z = X_f$ and $\contrf{Z}{\double\alpha} =
-\pi^*df$. Since these equations are global, it follows that $Z$ exists
globally and is unique. These two equations also guarantee that $Z$ belongs to
$\Sym(Y,\alpha)$.
\QED\enddemo
}  

\remark{Remark}
Since the fundamental vector fields associated to the right action of $\CA/D$
on $Y$ reproduce the Lie algebra elements when contracted with the connection,
it follows that they are infinitesimal symmetries. Moreover, one can deduce
from the local expression \recalt{loxexpreta} that they correspond to
the constant functions in $\Poisson$. In other words, the vector fields
$\eta_f$, with $f$ running through the constant functions in $\Poisson$,
generate the action of the structure group $\CA/D$ on the principal fiber
bundle $Y$. If $M$ is connected we thus can say that the kernel of the map
$\Poisson \to \HSym(M,\omega)$ corresponds to the (infinitesimal) action of the
structure group on $Y$.

\endremark

\proclaim{\thmm{symliftgroup}{Proposition}}
Let $G$ be a symmetry group of a connected \ssmfd/ $(M,\omega)$,
let
$\pi:{\Liealg h} \to \Liealg g$ be the central extension of the \glalg/ $\Liealg
g$ of
$G$ determined by $\omega$ \recalt{symdefcentralext}, and let $(Y,\alpha)$ be a
$D$-prequantum bundle over $M$ with projection $\pi^Y:Y \to M$. Then there
exists a momentum map for the
$G$-action if and only if there exists a Lie algebra morphism $H:\body
{\Liealg h} \to \Sym(Y,\alpha)$ compatible with the $G$-action, \ie,
each $H(V)$, $V\in \body {\Liealg h}$ projects to the fundamental vector
field of $\pi(V) \in \Liealg g$~:  $\pi^Y_*(H(V)) = (\pi (V))^M$. 

\endproclaim

\dontprint{
\demo{Proof}
\itemize
Suppose first that the map $H$ exists. From \recalt{infsymprop}(iii) we deduce
that for each $V\in \body\Liealg h$ there exists $f\in \Poisson$ such that
$H(V) = \eta_f$ and thus $(\pi (V))^M = X_f$, \ie, $(\pi (V))^M$ is globally
hamiltonian. Since the $\pi (V)$, $V\in \body \Liealg h$ generate $\Liealg g$,
we have proven that there exists a momentum map.

\itemize
Now suppose that there exists a momentum map $J$. Combining
\recalt{centralextbycocycle} and \recalt{cocyclebymomentum} we see that the
bracket on $\Liealg h = \Liealg g \times C$ is given by
$$
\lieb{(v,e)\,,\,(w,f)} = (\, \lieb{v,w} \,,\, \contrf{v,w}{\Omega_J} \,)
\mapob.
$$
We now define the map $\Hh:\body\Liealg h \to \Poisson$ by $\Hh(v,e) =
\contrs{v}{\double J} + e$. Computing $\Hh(\lieb{(v,e)\,,\,(w,f)})$ we find:
$$
\align
\Hh(\lieb{(v,e)\,,\,(w,f)})
&=
\contrs{\lieb{v,w}}{\double J} + \contrf{v,w}{\Omega_J}
\\&=
\PB{ \contrs{v}{\double J}, \contrs{w}{\double J}}
=
\PB{\Hh(v),\Hh(w)}
\mapob,
\endalign
$$
where the last equality follows from the fact that constant functions have zero
Poisson bracket. It follows that $\Hh$ is a Lie algebra morphism. Combining it
with the isomorphism $\Poisson \cong \Sym(Y,\alpha)$ we obtain the desired
result.
\QED\enddemo
} 

The results \recalt{infsymprop} and \recalt{symliftgroup} give us a motivation
for the introduction of a $D$-pre\-quantum bundle for a symplectic manifold. It
provides us with an injective representation of the Poisson algebra as vector
fields on $Y$, contrary to the representation by hamiltonian vector fields on
$M$ in which the (locally) constant functions disappear. Even better, it gives
us an isomorphism between the Poisson algebra and the infinitesimal symmetries
of the $D$-prequantum bundle. It also gives a nice interpretation of the
existence of a momentum map: it exists if and only if the infinitesimal action
of the central extension of $\Liealg g$ determined by the symplectic form can
be lifted to an infinitesimal action on the prequantum bundle $Y$.

\bigskip

In order to provide some motivation for prequantization in the sense of Kostant
and our weakened version in case the symplectic form is not even, we will use
the idea of quantization. Quantization is usually formulated a finding a
representation of the Poisson algebra on some space of functions with
additional requirements. In this paper we will forego the conditions concerning
a Hilbert space structure and irreducibility, nor will we discuss the
incompatibility between the various conditions. For this the interested reader
is referred to the existing literature (e.g., \cite{TW}, \cite{GGT} and
references cited therein). We will focus on a representation $\QZO$ of the
Poisson algebra $\Poisson$ on a space of functions $\QFS$ with the additional
condition that constant functions $r\cdot c_0 \in \Poisson$ ($r\in \RR$) are
represented by $r\cdot id(\QFS)$. To be more precise, we will look at even
linear maps
$\QZO: \Poisson \to \End(\QFS)$ satisfying the representation condition
$$
\lieb{\QZO(f),\QZO(g) } = -i\hbar\, \QZO(\PB{f,g}) 
\mapob,
$$
For a symplectic manifold
$(M,\omega)$ we have the obvious candidate $\QFS = \Ci(M, \CAC)$ of
$\CAC$-valued smooth functions on $M$ with $\QZO(f) = -i\hbar X_f$. But this
does not fulfill the addition condition because the hamiltonian vector fields of
constant functions are zero. 

Given a $D$-prequantum bundle $Y$, we can improve upon this situation by taking
$\QFS = \Ci(Y, \CAC)$ and $\QZO(f) = -i\hbar\, \eta_f$. Since $f\mapsto \eta_f$
is an injective representation, $\QZO$ is injective and no longer sends
constant functions to the zero operator. In order to see whether  $c_0
\in \Poisson$ ($r\in \RR$) is represented by $id(\QFS)$, we recall that
the vector field $\eta_{c_0}$ is the same as minus the fundamental vector field
associated to the action of $\CA_0/D$ on $Y$ (the minus sign comes from
$\contrf{\eta_{c_0}}{\double\alpha} = -\pi^*f = -1$). For
$\QZO(c_0)$ to act as the identity operator on a function $\phi\in \Ci(Y)$, the
function
$\phi$ should be of the form 
$$\phi(m,x+\xi) = \eexp^{-ix/\hbar} \phih(m,\xi)
\formula{locformphi}
$$ 
on a local
trivializing chart for the bundle $Y$. Since $x$ is a coordinate modulo $d$,
this implies that if we want the function $\phih$ to be non-zero, then
necessarily $d/\hbar \in 2\pi \ZZ$. If this is the case, the condition that
$\phi$ is of the form \recalf{locformphi} can be stated as the condition
$$
\phi(y\cdot  t) = \eexp^{-it/\hbar} \cdot \phi(y) \qquad y\in Y\ ,\ t\in
\CA_0/D
\mapob,
$$
where $y\cdot t$ denotes the right action of $t\in \CA_0/D$ on $Y$. This
description has the advantage that it is globally valid. Combining the local
expression \recalf{locformphi} with the local expression \recalf{loxexpreta} we
see that the subspace on which $\QZO(c_0)$ acts as the identity is invariant
under the action of the vector fields $\eta_f$. We thus obtain the following
result.

\proclaim{\thm{Proposition}}
There exists a non-zero subspace of $\Ci(Y, \CAC)$ on which $\QZO(c_0)$ acts as
the identity if and only if $D\subset 2\pi\hbar\ZZ$ (which implies but is not
equivalent to $\Per(\omega) \subset 2\pi\hbar\ZZ$). If this condition is
satisfied, the subspace in question is invariant under the representation
$\QZO$ and can be described as
$$
\{ \, \phi\in \Ci(Y, \CAC) \mid \forall a,y :
\phi(y\cdot a) = \eexp^{-ia/\hbar} \cdot \phi(y) \,\}
\mapob.
$$

\endproclaim

If $\CA_0/D$ were the whole structure group of the principal fiber bundle,
this would describe exactly the sections of an associated vector
bundle (associated to the representation $\rho(t\mod d) = \eexp^{it/\hbar}$ of
$\CA_0/D$ on $\CAC$). However, our structure group has the additional factor
$\CA_1$. Moreover, we have focussed attention on the functions $r\cdot c_0 \in
\Poisson$, which are constant functions in the Poisson algebra. But they do not
constitute all constant functions: a general constant function in $\Poisson$ is
of the form $r\cdot c_0 + s \cdot c_1$, $r,s\in \RR$. Since $c_1$ is odd and
$\QZO$ a representation and thus even, the result $\QZO(c_1)$, which is odd, can
never be a non-zero multiple of the identity operator, which is even. I now
arbitrarily decide that we want to represent $\QZO(c_1)$ as the zero operator.
In order to motivate this decision, note that \recalt{triviality}(ii) implies
that
$$
\Ci(Y,\CAC) \cong \Ci(\YO,\CAC) \oplus \Ci(\YO,\CAC) = \Ci(\YO,\CAC)^2
\mapob,
$$
in which the pair of functions $\phi^0,\phi^1 \in \Ci(\YO,\CAC)$ corresponds to
the function ${\phi\in \Ci(Y,\CAC)}$ given by
$$
\phi(y,\xi) = \phi^0(y) + \xi \cdot \phi^1(y) \qquad y\in \YO
\mapob.
$$
Moreover, it is immediate from \recalf{loxexpreta} that the first factor
$\Ci(\YO,\CAC) \cong \{\,(\phi,0) \mid \phi\in \Ci(\YO,\CAC) \,\} \subset
\Ci(Y,\CAC)$ of functions independent of $\xi$ is invariant under all
operators $\QZO(f)$, $f\in \Poisson$. This is also the largest subspace on
which $\QZO(c_1)$ acts as the zero operator because $\QZO(c_1) = i\hbar
\partial_\xi$ maps $(\phi^0,\phi^1)$ to $(i\hbar \phi^1, 0)$. We thus see
that $\Ci(Y,\CAC)$ naturally splits as a direct sum of two copies of
$\Ci(\YO,\CAC)$, one of which is the null space of $\QZO(c_1)$ which is
invariant under the action of $\Poisson$. If we think of quantization as a
quest for an irreducible representation, the choice to look at this null space
of $\QZO(c_1)$ becomes natural. 

Since this null space is the space of functions independent of the global
coordinate $\xi$, we can also describe it as
$$
\{\, \phi\in \Ci(Y,\CAC) \mid \forall\tau,y : \phi(y\cdot
\tau) = \phi(y) \,\}
\mapob,
$$
where $y\cdot \tau$ denotes the right action of $\CA_1 \subset \CA/D$ on $Y$. 
If we are interested in the intersection of this null space with the space on
which $\QZO(c_0)$ acts as the identity, we can combine the two descriptions to
obtain the function space $\QFS = {\{\QZO(c_0) = id\} \cap \{\QZO(c_1) = 0\}}$~:
$$
\QFS = \{\, \phi \in \Ci(Y,\CAC) \mid \forall \cadb t,y : 
\phi(y \cdot \cadb t) = \eexp^{-it/\hbar}\phi(y) \,\}
\mapob.
$$
From this we see that $\QFS$ is the set of smooth sections of the vector bundle
$L$ with typical fiber $\CAC$ associated to the principal fiber bundle $Y$ by
the representation $\rho: \CA/D \to \End(\CAC)$, $\rho(\cadb t) =
\eexp^{it/\hbar}$.  And as said before, this representation makes sense only
if $D \subset 2\pi \hbar \ZZ$. Since $Y$ has a connection $\double\alpha$, we
have an induced connection $\nabla$ on $L$. Computing the curvature of
$\nabla$, we find
$$
\curv(\nabla) = \rho_*\curv(\double\alpha) = i\omega_0 /\hbar
\mapob.
$$
We see that this curvature is independent of the choice of $D$. But of course
we have to choose $D$ such that $\Per(\omega) \subset  D \subset
2\pi\hbar \ZZ$. Choosing $D = 2\pi\hbar \ZZ$ imposes the least number of
restrictions on $\Per(\omega)$. We can summarize this as follows.

\proclaim{\thm{Proposition}}
There exists a non-zero subspace $\QFS$ of $\Ci(Y, \CAC)$ on which $\QZO(c_0)$
acts as the identity and on which $\QZO(c_1)$ acts as the zero operator
if and only if $\Per(\omega) \subset 2\pi\hbar\ZZ$. 

If this
condition is satisfied, $\QFS$ is invariant under the
representation $\QZO$ and can be described as
$$
\QFS = 
\{ \, \phi\in \Ci(Y, \CAC) \mid \forall\ \cadb t \in \CA/D,y\in Y :
\phi(y\cdot \cadb t) = \eexp^{-it/\hbar} \cdot \phi(y) \,\}
\mapob,
$$
\ie, $\QFS$ is the set of smooth sections of the complex line bundle
$L$ over $M$ associated to $Y$ by the representation $\rho(\cadb t) =
\eexp^{it/\hbar}$. It can also be described as 
$$
\QFS = 
\{ \, \phi\in \Ci(\YO, \CAC) \mid \forall\ t \in \CA_0/D,y\in \YO :
\phi(y\cdot  t) = \eexp^{-it/\hbar} \cdot \phi(y) \,\}
\mapob.
$$
Said differently, $L$ is the complex line bundle
over $M$ associated to $\YO$ by the representation $\rho( t) =
\eexp^{it/\hbar}$ of $\CA_0/D$ on $\CAC$. 
The induced linear connection $\nabla$ on $L$ has curvature
$i\omega_0/\hbar$. In other words, $L$ is the prequantization line bundle in
the sense of Kostant. In terms of this line bundle the representation $\QZO$
takes the form
$$
\QZO(f) s = -i\hbar \nabla_{X_f} s + f^0s
\mapob,
$$ 
for $f\in \Poisson$ and $s$ a smooth section of $L$.

\endproclaim

\dontprint{
\demo{Proof}
The only thing that remains to be proven is the expression for $\QZO(f)$ in
terms of the line bundle. In the equivalence between sections of a line bundle
and functions on the principal fiber bundle, the action of $\nabla_X$ on a
section corresponds to the action of $\Xt$ on the corresponding function with
$\Xt$ being the horizontal lift of $X$. It is immediate from
\recalf{locexpralpha} that $\Xt_f$ is (locally) given by 
$$
\Xt_f = X_f -\contrf{X_f}{(\theta_i)_0} \fracp{}{x_i}
-\contrf{X_f}{(\theta_i)_1} \fracp{}{\xi_i}
\mapob,
$$
and thus
$$
\eta_f = \Xt_f - f^0  \fracp{}{x_i}
-  f^1  \fracp{}{\xi_i}
\mapob.
$$
Applying $-i\hbar \eta_f$ to an element of $\QFS$ in the local form
\recalf{locformphi} and independent of $\xi_i$ is the same as applying $-i\hbar
\Xt_f$ plus multiplication by $f^0 $. 
\QED\enddemo
} 

To summarize: we started with a $D$-prequantum bundle and we looked for a
representation of the Poisson algebra satisfying certain conditions. Such a
quest is quite natural in quantization. We ended up with the line bundle of
prequantization in the sense of Kostant in its weaker form. The only
unjustified point in our argument was the condition to look at the subspace on
which $\QZO(c_1)$ acts as the zero operator. Future research in unitary
representations of \glgrp/s and in quantization of \ssmfd/s should determine
whether this ad hoc assumption is the right thing to do. Note however
that in the ungraded case (no $\CA_1$, no $c_1$) it is the subspace $\QFS$
which is the starting point of the geometric quantization procedure\slash orbit
method.

\bigskip

To end, let us look at the special cases in which the symplectic form is
homogeneous (and the manifold connected). If $\omega$ is even, we know that the
Poisson algebra is (isomorphic to) $\Ci(M) \oplus \RR$; more precisely, $f =
f^0 c_0 + f^1 c_1 :M\to C$, $f^\alpha \in \Ci(M)$ belongs to $\Poisson$ if
and only if $f^1$ is constant (and thus real). Moreover, the subset $\Ci(M)
\cong \Ci(M) \oplus \nul \subset \Poisson$ is a subalgebra. For this subalgebra
the definition of hamiltonian vector field and Poisson bracket is exactly the
classical ungraded definition: $\omega$ is non-degenerate,
$\contrf{X_f}{\omega} = df$ and $\PB{f,g} = X_fg$. Any prequantum bundle $Y$ is
of the form $\YO
\times \CA_1$ with a trivial connection $d\xi \otimes c_1$ on the $\CA_1$ part
\recalt{triviality}. And the representation $\QZO$ on $\QFS$ is an injective
representation of $\Ci(M) \subset \Poisson$ (it kills the additional part of
functions $r \cdot c_1$, $r\in \RR$). We see that in the even case we can
completely forget about the $c_1$ part, it does not contain any relevant
information. We also have equivalence between prequantization according to
Souriau and Kostant. And when we forget about the $c_1$ part, the Poisson
algebra becomes just $\Ci(M)$, of which $\QZO$ is an injective representation.

If $\omega$ is odd, the Poisson algebra is again (isomorphic to) $\RR \oplus
\Ci(M)$, but this time we have $f = f^0 c_0 + f^1 c_1 \in \Ci(M,C)
$ belongs to $\Poisson$ if and only if $f^0$
is constant (and thus real). As in the even case, the subset $\Ci(M)
\cong \nul \oplus \Ci(M) \subset \Poisson$ is a subalgebra. For this
subalgebra the definition of hamiltonian vector field and Poisson bracket is
again exactly the classical ungraded definition: $\omega$ is non-degenerate,
$\contrf{X_f}{\omega} = df$ and $\PB{f,g} = X_fg$. The difference with the even
case lies in parity. The even part $\Poisson_0$ of the Poisson algebra is a
subalgebra. Hence the intersection of $\Poisson_0$ with the previously defined
subalgebra $\Ci(M)$ is also a subalgebra. But since $c_1$ is odd we have
$\Poisson_0 = \RR \oplus \Ci(M)_1$ and $\Poisson_0 \cap \Ci(M) = \Ci(M)_1$.
Any prequantum bundle $Y$ is of the form $\YO
\times \CA_1$, but now with $(\YO, \alpha_0)$ a principal $\CA_0/D$-fiber bundle
with connection whose curvature is zero (one could choose the trivial bundle
$\YO = M \times \CA_0/D$ and $\alpha_0 = dx \otimes c_0$), and a non trivial
connection
$(\theta_1 + d\xi) \otimes c_1$ on the $\CA_1$ part \recalt{triviality}. The
representation $\QZO$ kills the constant functions in $\Ci(M)$, but it
represents the constant functions $r\cdot c_0$, $r\in \RR$ as $r\cdot id_\QFS$.
The restriction of $\QZO$ to $\Ci(M)$ is not injective, but its restriction to
$\Poisson_0$ (and thus a fortiori $\Ci(M)_1$) is.

\enddocument